\documentclass[trackchanges]{aastex701}

\usepackage{amsmath}
\usepackage{amsfonts}

\usepackage{xcolor}

\begin{document}

\title{A Pilot Study of Mildly Recycled Pulsars: A Case Study of PSR~J2338+4818}

\author[orcid=0009-0005-5141-8305]{Yujie Chen}
\affiliation{College of Physics, Guizhou University, Guiyang 550025, P.R. China}
\email{gs.yujiechen23@gzu.edu.cn}

\author[orcid=0009-0001-6693-7555]{Yujie Lian}
\affiliation{Institute for Frontiers in Astronomy and Astrophysics, Beijing Normal University, Beijing 102206, P.R. China}
\affiliation{School of Physics and Astronomy, Beijing Normal University, Beijing 100875, P.R. China}
\email{yujielian@mail.bnu.edu.cn}

\author[orcid=0000-0002-5344-9552]{Yujie Wang}
\affiliation{Shanghai Astronomical Observatory, Chinese Academy of Sciences, 80 Nandan Road, Shanghai 200030, P.R. China}
\affiliation{School of Astronomy and Space Sciences, University of Chinese Academy of Sciences, Beijing 100049, China}
\email{wangyj@shao.ac.cn}

\author[orcid=0000-0002-2394-9521]{Liyun Zhang}
\affiliation{College of Physics, Guizhou University, Guiyang 550025, P.R. China}
\affiliation{International Centre of Supernovae, Yunnan Key Laboratory, Kunming 650216, P.R. China}
\email{liy\_zhang@hotmail.com}
\correspondingauthor{Liyun Zhang, liy\_zhang@hotmail.com}

\author[orcid=0000-0003-0597-0957]{Lei Qian}
\affiliation{Guizhou Radio Astronomical Observatory, Guizhou University, Guiyang 550025, P.R. China}
\affiliation{National Astronomical Observatories, Chinese Academy of Sciences, 20A Datun Road, Chaoyang District, Beijing 100101, P.R. China}
\affiliation{College of Astronomy and Space Sciences, University of Chinese Academy of Sciences, Beijing 100049, P.R. China}
\email{lqian@nao.cas.cn}
\correspondingauthor{Lei Qian, lqian@nao.cas.cn}

\author[orcid=0000-0001-9754-9777]{Zhichen Pan}
\affiliation{Guizhou Radio Astronomical Observatory, Guizhou University, Guiyang 550025, P.R. China}
\affiliation{National Astronomical Observatories, Chinese Academy of Sciences, 20A Datun Road, Chaoyang District, Beijing 100101, P.R. China}
\affiliation{College of Astronomy and Space Sciences, University of Chinese Academy of Sciences, Beijing 100049, P.R. China}
\email{panzc@nao.cas.cn}
\correspondingauthor{Zhichen Pan, panzc@nao.cas.cn}

\author[orcid=0000-0002-8870-981X]{Shuo Cao}
\affiliation{Institute for Frontiers in Astronomy and Astrophysics, Beijing Normal University, Beijing 102206, P.R. China}
\affiliation{School of Physics and Astronomy, Beijing Normal University, Beijing 100875, P.R. China}
\email{caoshuo@bnu.edu.cn}

\author[orcid=0000-0001-6051-3420]{Dejiang Yin}
\affiliation{College of Physics, Guizhou University, Guiyang 550025, P.R. China}
\email{gs.djyin21@gzu.edu.cn}

\author[orcid=0009-0008-4109-744X]{Baoda Li}
\affiliation{College of Physics, Guizhou University, Guiyang 550025, P.R. China}
\email{gs.bdli21@gzu.edu.cn}

\author{Ruili He}
\affiliation{College of Physics, Guizhou University, Guiyang 550025, P.R. China}
\email{1778968351@qq.com}

\author{Tong Liu}
\affiliation{National Astronomical Observatories, Chinese Academy of Sciences, 20A Datun Road, Chaoyang District, Beijing 100101, P.R. China}
\email{tongliu@bao.ac.cn}

\author{Wenze Li}
\affiliation{College of Physics, Guizhou University, Guiyang 550025, P.R. China}
\email{wenze_li@qq.com}

\author{Yichi Zhang}
\affiliation{National Astronomical Observatories, Chinese Academy of Sciences, 20A Datun Road, Chaoyang District, Beijing 100101, P.R. China}
\affiliation{School of Astronomy and Space Sciences, University of Chinese Academy of Sciences, Beijing 100049, China}
\email{zhangyichi25@mails.ucas.ac.cn}

\author{Yifeng Li}
\affiliation{National Time Service Center, Chinese Academy of Sciences, Xi'an 710600, P.R. China}
\affiliation{University of Chinese Academy of Sciences, Beijing 100049, P.R. China}
\email{liyifeng@ntsc.ac.cn}

\author{Qiaoli Hao}
\affiliation{National Astronomical Observatories, Chinese Academy of Sciences, 20A Datun Road, Chaoyang District, Beijing 100101, P.R. China}
\email{qlhao@nao.cas.cn}

\author{Jinyou Song}
\affiliation{National Astronomical Observatories, Chinese Academy of Sciences, 20A Datun Road, Chaoyang District, Beijing 100101, P.R. China}
\email{songjinyou@nao.cas.cn}

\author{Shuangyuan Chen}
\affiliation{National Astronomical Observatories, Chinese Academy of Sciences, 20A Datun Road, Chaoyang District, Beijing 100101, P.R. China}
\email{sychen@nao.cas.cn}

\author{Xingyi Wang}
\affiliation{National Astronomical Observatories, Chinese Academy of Sciences, 20A Datun Road, Chaoyang District, Beijing 100101, P.R. China}
\email{wangxingyi@nao.cas.cn}

\author{Xianghua Niu}
\affiliation{National Astronomical Observatories, Chinese Academy of Sciences, 20A Datun Road, Chaoyang District, Beijing 100101, P.R. China}
\email{xhniu@nao.cas.cn}

\author{Minglei Guo}
\affiliation{National Astronomical Observatories, Chinese Academy of Sciences, 20A Datun Road, Chaoyang District, Beijing 100101, P.R. China}
\email{guominglei@nao.cas.cn}

\author{Menglin Huang}
\affiliation{National Astronomical Observatories, Chinese Academy of Sciences, 20A Datun Road, Chaoyang District, Beijing 100101, P.R. China}
\email{huangmenglin@nao.cas.cn}

\begin{abstract}
Mildly recycled pulsars are neutron stars partially spun up through relatively short mass-transfer phases, typically with massive carbon–oxygen (CO) or oxygen–neon–magnesium (ONeMg) white dwarf companions.
PSR~J2338+4818, a mildly recycled pulsar, was discovered with the Five-hundred-meter Aperture Spherical Telescope (FAST).
As a pilot study on the formation and evolutionary pathways of mildly recycled pulsars, 
we present the updated timing solution for PSR~J2338+4818 and examine its single pulses and scintillation properties.
Aided by the sensitivity of FAST, the single pulses of PSR~J2338+4818 were systematically studied.
27,228 single pulses with S/N $> 7$ have been detected in our observations.
For the FAST ultra-wideband observation on MJD~61045, the receiver was still in the technical commissioning phase,
and then only a preliminary single-pulse search was performed.
Pulse nulling was examined using a Markov Chain Monte Carlo (MCMC) method, but no evidence for nulling was found.
The possible long-term nulling reported by previous studies did not occur in any of our observations in either the 1.0 to 1.5~GHz band or the 300 to 600~MHz band.
Interstellar scintillation is evident in our observations.
The measured scintillation timescales and bandwidths range from 2.93 to 25.26~minutes and 1.68 to 27.41~MHz, respectively.
In all observations, no clear scintillation arc was found in the secondary spectra of PSR~J2338+4818.
\end{abstract}



\section{Introduction}\label{sec:intro}
More than 4,000 pulsars have been detected in the Milky Way to date, among which over 500 are recycled pulsars in binary systems. 
Following the supernova explosion of the primary star that forms a neutron star (NS), if the system remains bound, the NS can accrete mass and angular momentum as the secondary companion evolves and fills its Roche lobe. 
This process is referred to as pulsar recycling, during which the neutron star regains radio emission, is spun up to shorter periods, and exhibits a significantly reduced surface magnetic field ($B \lesssim 10^{10}\,\mathrm{G}$)~\citep{1991PhR...203....1B}.
Pulsars that have undergone this process are termed recycled pulsars.
Systems containing a recycled pulsar and a white dwarf (WD) are widely regarded as the evolutionary endpoints of low-mass and intermediate-mass X-ray binaries (LMXBs and IMXBs; \citealt{2023pbse.book.....T}).
A key observable for understanding the formation of recycled pulsars is their spin period, which is closely related to the total amount of mass accreted by the neutron star \citep{2012MNRAS.425.1601T}. 
The accreted mass, in turn, depends primarily on the mass-transfer timescale of the progenitor X-ray binary, as well as on the mode of mass transfer and mass loss from the system. 
Since the mass-transfer timescale is determined by the evolutionary state and initial mass of the donor star, the efficiency of the recycling process and the nature of the final system depend heavily on the donor’s zero-age main sequence mass. 
Generally, low-mass companions ($\lesssim 2.3 \, M_\odot$) lead to low-mass x-ray binary phases, producing fully recycled millisecond pulsars (MSPs) with $P < 10~{\rm ms}$ and light helium white-dwarf (He-WD) companions \citep{2006csxs.book..623T}.
For companions with intermediate initial masses (approximately $2.3 - 8 \, M_\odot$), the mass-transfer stage is relatively brief due to the rapid evolution of the donor star. 
In such cases, the NS is only partially spun up, resulting in mildly recycling. 
These mildly recycled pulsars are characterized by spin periods typically between 10 and 200~ms \citep{2006csxs.book..623T}. 
The companions in these binary systems are typically carbon–oxygen (CO) or oxygen–neon–magnesium (ONeMg) white dwarfs in low-eccentricity ($e \lesssim 10^{-3}$) orbits \citep{2000ApJ...530L..93T,2011MNRAS.416.2130T,2012MNRAS.425.1601T}.
In systems originating from high-mass X-ray binaries (HMXBs), the companion can be another neutron star, forming a double neutron stars (DNS) system. 
A example is the DNS system PSR~J1930-1852, which contains a mildly recycled pulsar with a spin period of 185~ms and an orbital period of 45~days \citep{2015ApJ...805..156S}.
Orbital eccentricities are typically large in these cases owing to the second supernova explosion, with most systems having $e \gtrsim 0.1$, while a small population of DNS systems shows lower eccentricities, such as PSR J1946+2052 ($e \approx 0.064$) and PSR J1325$-$6253 ($e \approx 0.064$) \citep{2018ApJ...854L..22S,2022MNRAS.512.5782S}.


To systematically study this population of the mildly recycled pulsars, we planned a survey by using the Five-hundred-meter Aperture Spherical radio Telescope (FAST). 
The observable sky of FAST covers a declination range of $-14.4^\circ \leq \delta \leq +65.6^\circ$ \citep{2020Innov...100053Q}.
More than 40 mildly recycled pulsars within the FAST-detectable sky were selected.
Among these pulsars, PSR~J2338+4818 has been observed by FAST for a long time, since it is the first pulsar discovered by FAST. 
For this reason, we study PSR~J2338+4818 as a pilot study for the survey of the mildly recycled pulsars.

PSR~J2338+4818 was discovered and confirmed by the Five-hundred-meter Aperture Spherical radio Telescope (FAST; \citealt{Nan2011FAST}) on August 9$^{\rm th}$, 2017, 
from the drift scan data taken on August 4$^{\rm th}$ and 6$^{\rm th}$.
It was designated (and hereafter) as PSR~J2338+4818 which is the first candidate (and then confirmed as a new pulsar) from FAST.
It is with a 118-ms spin period, and a dispersion measure (${\rm DM}$) of $35.3 {\ \rm cm^{-3}\,pc}$. 
It was then observed for two years with the Effelsberg telescope \citep{2021MNRAS.508..300C}.
Its timing solution was derived by using \textsc{DRACULA}\footnote{\url{https://github.com/pfreire163/Dracula}} \citep{Freire2018DRACULA}.
It is an old (with an inferred age of $\sim 0.95{\ \rm Gyr}$) mildly recycled pulsar,
locating in a binary system, 
with an orbital period of about 95~days, 
and a mild eccentricity of 0.0018237(9). 
\citet{2021MNRAS.508..300C} report that the minimum mass of the companion is 1.049~$\rm M_{\odot}$ (the mass function of the system is 0.19263(4) and assuming a pulsar mass of 1.4~$M_{\odot}$).
The companion is mostly likely a carbon–oxygen white-dwarf \citep{1984ApJS...54..335I,2011MNRAS.416.2130T}.

A possible long-term nulling of PSR~J2338+4818 was observed,
during which the pulsar was not detected in some observations lasting a few hours \citep{2021MNRAS.508..300C}.
It is interesting that nulling has been observed in some pulsars, and studies suggest that the nulling fraction does not show any obvious correlation with age or other intrinsic parameters \citep{2019JApA...40...42K}.
The possible nulling in PSR~J2338+4818, an old mildly recycled pulsar, if confirmed, adds to the sample of pulsars exhibiting this behavior, helping to explore the underlying mechanisms behind the phenomenon.
The PSR~J2338+4818 shows pronounced interstellar scintillation and kept a high signal-to-noise ratio (${\rm S/N}$) in the FAST observations.

In this paper, we present an in-depth study of PSR~J2338+4818 with FAST.
The observations and data reduction are described in Section~\ref{sec:2}.
The results are presented in Section~\ref{sec:3}, which includes the timing analysis (Section~\ref{subsec:4}), 
the single-pulse and nulling analysis (Section~\ref{subsec:5}).
and the scintillation analysis (Section~\ref{subsec:6}).
The discussion and conclusions are presented in Section~\ref{Conclusions}.

\section{Observation and Data Reduction}\label{sec:2}
To redetect PSR~J2338+4818 with FAST and as is was discovered in a subband (290 to 340~MHz) from the ultra-wideband receiver of FAST covering  270 to 1620~MHz, one observation was done with a 300 to 600~MHz testing receiver of FAST, in October 6$^{\rm th}$, 2023, lasting for 1320~s, 
after the other was done with the L-band multibeam receiver.
The PSR~J2338+4818 was detected in both observations.
Since then, the timing campaign was started till February, 2026.

Among all the 36~observations, the central beam of FAST's 19-beam L-band receiver was used for 34~observations.
The frequency range is 1.0 to 1.5~{\rm GHz}, with 4096~channels (corresponding to a channel width of 0.122~{\rm MHz}).
The sampling time for L-band receiver is 49.152~$\mu\mathrm{s}$.
The typical system temperature was about 22~{\rm K} at 1250~{\rm MHz} \citep{2020RAA....20...64J}. 
For the 300 to 600~{\rm MHz} receiver, 
the digital backend of this testing receiver record date with a frequency range of 0 to 625~{\rm MHz},
while we used the data of 300 to 600~{\rm MHz} (31465~channels with a 9.535~{\rm kHz} channel width). 
The sampling time is $838.861\,\mu\mathrm{s}$.
For the ultra-wideband receiver, the effective frequency coverage was 500--3330~{\rm MHz}, and the backend recorded the data in four sub-bands with a channel width of 0.26855~{\rm MHz} (see~\citealt{2023RAA....23g5016Z}).
The sampling time for ultra-wideband receiver is $100.543\,\mu\mathrm{s}$.

For the 2026-01-05 ultra-wideband observation, the effective frequency coverage was 500–3300 MHz, and the receiver recorded the data in four sub-bands \citep{2023RAA....23g5016Z}.

The detailed information, including observing dates, lengths, the ${\rm S/N}$ values of PSR~J2338+4818 and numbers of polarizations can be found in Table~\ref{table:FAST_observations}.
\begin{table}
\centering 
\caption{FAST Observations of J2338+4818}
\begin{tabular*}{\textwidth}{@{\extracolsep{\fill}}cccccccc@{}}
\hline
Observation date & MJD & Length (s) & ${\rm S/N}$ & $N_{\mathrm{pol}}$ & Total pulses  &  Pulses ($\mathrm{S/N} > 7$) &NF(\%) \\
\hline\hline
2023-09-04 & 60190 & 3600 & 293 & 4  & 29823  & 353 & 0.2~\% \\
2023-10-06$^{a}$ & 60222 & 1320 & 150 & 2 & 10116  & 136 &  4.0~\% \\
2023-10-09 & 60225 & 1200 & 219 & 4 & 9612  & 20 & 0.4~\% \\
2023-11-05 & 60253 & 3600 & 154 & 4 & 29818   & 0 &  0.2~\% \\
2023-12-11 & 60289 & 7800 & 210 & 4 & 64551   & 1 & 0.5~\% \\
2023-12-23 & 60301 & 3300 & 143 & 4 & 9851   & 0 & 18.0~\%\\
2024-01-11 & 60320 & 7200 & 271 & 4 & 60156   & 0 & 1.7~\% \\
2024-02-07 & 60347 & 7200 & 65 & 4 & 59803   & 0  & 36.3~\% \\
2024-03-17 & 60386 & 11860 & 284 & 2 & 99852 & 2132  & 0.6~\% \\
2024-04-16 & 60416 & 4365  & 251 & 4 & 36224 & 52  &  0.3~\%\\
2024-05-29 & 60459 & 4560  & 137 & 4 & 37988  & 0  &  1.1~\%\\
2024-08-06 & 60527 & 7200 & 179 & 2 & 60792  & 0 & 0.3~\%  \\
2024-09-01 & 60553 & 1200 & 182 & 4 & 9001  & 1114 & $\sim~0$~\% \\
2024-09-06 & 60558 & 1200 & 120 & 4 & 9001  & 0 & 0.5~\% \\
2024-09-08 & 60560 & 1200 & 143 & 4 & 9001  & 0 & 0.7~\% \\
2024-09-13 & 60565 & 1200 & 109 & 4 & 9001  & 0 & 5.9~\% \\
2024-09-29 & 60581 & 1200 & 113 & 4 & 9001  & 1 & 3.7~\% \\
2024-11-06 & 60620 & 1800 & 180 & 4 & 15298  & 133 & 0.3~\% \\
2024-12-04 & 60648 & 1800 & 183 & 4 & 15296   &  6 & 0.5~\% \\
2024-12-14 & 60658 & 3600 & 219 & 2 & 29825  & 1686 & $\sim~0$~\% \\
2024-12-22 & 60666 & 3600 & 211 & 2 & 29826  & 8899 & $\sim~0$~\% \\
2024-12-23 & 60667 & 3600 & 202 & 2 & 29826 & 4492 & $\sim~0$~\% \\
2025-03-09 & 60743 & 1800 & 178 & 4 & 15296  & 367  & 19.4~\% \\
2025-03-10 & 60744 & 10800 & 168 & 2 & 90301 & 0    & 0.3~\%\\
2025-03-21 & 60755 & 7200 & 164 & 4 & 59400  & 4641 & $\sim~0$~\%  \\
2025-03-23 & 60757 & 7200 & 198 & 4 & 60131 & 132    & 0.1~\% \\
2025-03-24 & 60758 & 7200 & 197 & 4 & 60145  & 17   & 1.2~\% \\
2025-04-09 & 60774 & 3600 & 178 & 4 & 30461 & 68    & 3.4~\% \\
2025-04-19 & 60784 & 8400 & 108 & 2 & 53899 & 0   & 8.2~\% \\
2025-09-03 & 60920 & 3600 & 130 & 4 & 29941  & 61   & 10.3~\% \\
2025-09-04 & 60921 & 3600 & 104 & 4 & 29947 & 42   & 12.5~\%\\
2025-09-05 & 60922 & 3600 & 84 & 4 & 29906 & 42   &  5.9~\%  \\
2025-12-10 & 61019 & 16920 & 202 & 2 & 140860 & 705 & $\sim~0$~\% \\
2026-01-02 & 61042 & 14400 & 241 & 2 & 111700 & 736 & 0.1~\% \\
2026-01-05$^{b}$ & 61045 & 12000 & 40 & 2 & 101180 & 1248 & -- \\
2026-02-19 & 61090 & 11595 & 223 & 2 & 92179 & 144 &  0.1~\% \\
\hline\hline
\end{tabular*}
\label{table:FAST_observations}
\begin{flushleft}
Note: $^{a}$ {The 300--600~{\rm MHz} observation.} $^{b}$ {The ultra-wideband observation.}
Most data were taken with the 19-beam L-band receiver (1--1.5~{\rm GHz}, 4096 channels, 0.122~{\rm MHz} channel width, $49.152\,\mu\mathrm{s}$ sampling), 
while the observation on 2023-10-06 was taken with the low frequency receiver (300--600~{\rm MHz}, 31465 channels, 9.535~{\rm kHz} channel width, $838.861\,\mu\mathrm{s}$ sampling).
The last three columns show the total number of pulses, the number of pulses detected with ${\rm S/N} > 7$, and the estimated nulling fraction (NF) for each observation, respectively.
As the FAST ultra-wideband receiver was still in the technical commissioning phase, we performed only a preliminary single-pulse search with \textsc{TransientX} on the ultra-wideband observation (MJD~61045).
\end{flushleft}
\end{table}

To prepare for the timing data, 
we used \textsc{DSPSR} \citep{2011PASA...28....1V} for folding, 
and \textsc{PSRCHIVE} \citep{2012AR&T....9..237V} to post-process the resulting archive files. 
In processing PSRFITS files with \textsc{DSPSR}, 
we set the number of phase bins in the pulse profile to 128 (\texttt{-b} option),
folded the data with the ephemeris from the \textsc{psrcat} software (\texttt{-E} option), 
created 6~${\rm s}$ sub-integrations (\texttt{-L} option), 
and combined multiple sub-integrations into a single archive file(\texttt{-A} option). 
Within \textsc{PSRCHIVE}, we employed the \texttt{pazi} routine to remove RFIs in both the time and frequency domains. 
The \texttt{pam} routine was used to perform merging operations along the time, frequency, and polarization on the interference-corrected archive files.
For each observation, the total integration time was divided into 16 equal sub-intervals.
The integration time per TOA varied from approximately 75 to 1057~s, depending on the duration of each observation.
We generated a standard template using the \texttt{paas} routine and used the \texttt{pat} routine to extract times of arrivals (TOAs).
The low frequency data was only used for pulse profile comparison.
Due to the changing of cables of the ultra-wideband and 300–600~{\rm MHz} receivers during commissioning, the time offsets of the receivers were not determined for our observations.
Thus, we do not use the ultra-wideband and 300 to 600~{\rm MHz} data for timing.

To prepare scintillation data, 
we used the \texttt{psrflux} routine from \textsc{PSRCHIVE} to generate the dynamic spectrum from the archive files. 
The sub-integration time and frequency channel resolution of each archive are 6~${\rm s}$ and 0.98~{\rm MHz}, respectively.
The scintillation timescale and bandwidth can be obtained from the auto-correlation function (ACF) of the dynamic spectrum. 
The ACF, namely C ({$\Delta \nu$}, {$\Delta \tau$ }), is described as \citep{1990ARA&A..28..561R}:
\begin{equation}
C(\Delta\nu, \Delta \tau) = \sum_\nu \sum_t \Delta D(\nu, t) \Delta D(\nu + \Delta\nu, t + \Delta \tau),
    \label{eq:1}
\end{equation}
where $\Delta\nu$ and $\Delta \tau$ are the frequency and time lag, respectively, $\Delta D(\nu, t)$ is the pulsar intensity as a function of frequency $\nu$ and time $t$. 
The deviation of the dynamic spectrum from its mean is defined as:
\begin{equation}
\Delta D(\nu, t) = D(\nu, t) - \overline{D},
\label{eq:2}
\end{equation}
where the $\overline{D}$ indicates averaging over time and frequency. 
The normalized ACF, namely N($\Delta\nu, \Delta t$), is expressed as:
\begin{equation}
N(\Delta\nu, \Delta t) = \frac{C(\Delta\nu, \Delta t)}{C(0,0)},
\label{eq:3}
\end{equation}
In this work, the ACF is calculated with the \textsc{Scintools} package\footnote{\url{https://github.com/danielreardon/scintools}}, following the approach of \citet{2020ascl.soft11019R}. 
The secondary spectrum is obtained from the two-dimensional Fourier transform of the dynamic spectrum (see details in e.g., \cite{2020ascl.soft11019R}):
\begin{equation}
P(f_t, f_\lambda) = 10 \log_{10} \bigl( |\widetilde{D}(t, \lambda)|^2 \bigr),
\label{eq:4}
\end{equation}
where $\widetilde{D}(t, \lambda)$ denotes the two-dimensional Fourier transform of $\Delta D$, and $f_t$ and $f_\lambda$ represent the conjugate time and wavelength, respectively.

The polarisation calibration was realised with the noise signal by using the \texttt{pac} routine of the \textsc{PSRCHIVE} on the only one observation (on~2024-09-01, MJD~60553) with noise injection.
We did not perform flux calibration since our observation was not absolute flux-calibrated using a noise generator. 
We used the \texttt{rmfit} routine to fit the Faraday rotation measure (RM).

To investigate the single-pulse and possible pulse nulling, 
we generated single pulses for all observations using \textsc{DSPSR} in single-pulse mode (\texttt{-s} option). 
For all observations except the 300–600~{\rm MHz} data (MJD~60222), we used 1024~bins.
For the 300–600~{\rm MHz} data, 64~bins were chosen to match the sampling period and bin width, ensuring without over-sampling.
Since the FAST ultra-wideband receiver is still in the technical commissioning phase, we performed only a preliminary single-pulse search on the ultra-wideband observation (MJD~61045) using \textsc{TransientX}\footnote{\url{https://github.com/ypmen/TransientX}} \citep{2024A&A...683A.183M}.
Pulses from different epochs were phase-aligned and combined using the \texttt{psradd} routine in \textsc{PSRCHIVE} after radio-frequency interference mitigation. 
The data were then averaged over frequency channels and polarization to obtain total-intensity single-pulse sequences as a function of rotational phase and generated single-pulse stacks.

Estimating the nulling properties requires defining the ON-pulse and OFF-pulse phase windows.
The OFF-pulse window is expected to contain only radiometer noise,
while the ON-pulse window is defined as the phase range where the signal exceeds three times the root-mean-square (RMS) of the OFF-pulse windows.
We first formed the integrated pulse profile and selected ON-pulse and OFF-pulse windows of equal width as shown in Figure~\ref{fig:residual_profile}. 
Histograms of the ON-pulse and OFF-pulse energies were constructed to characterise the energy distributions. 
The OFF-pulse distribution is expected to be well described by Gaussian noise, whereas the ON-pulse distribution may contain contributions from both null and emission states. 
In the presence of pulse nulling, the ON-pulse histogram would show an excess of samples at energy levels consistent with the OFF-pulse noise distribution, representing the null component. 
The remaining part of the ON-pulse distribution corresponds to the emission component, which may itself be composed of one or more sub-populations. 
The observed ON-pulse energy distribution can therefore be regarded as the sum of the null and emission components.

\begin{table}
\centering
\caption{Timing solution of J2338+4818} 
\begin{tabular}{lc}
\hline
Pulsar name                                              & {J2338+4818}         \\
\hline
Parameters                                                         & {}                    \\
Right Ascension, $\alpha$ (J2000)                         \dotfill & {23:38:06.21079(5)}     \\  
Declination, $\delta$ (J2000)                             \dotfill & {+48:18:31.8253(6)}     \\  
Spin Frequency, $F$ ($\mathrm{s^{-1}}$)                   \dotfill & {8.4238723393473(4)}    \\
Spin Frequency Derivative, $\dot{F}$ ($\mathrm{s^{-2}}$)  \dotfill & {-1.6361(2) $\times 10^{-16}$}  \\  
Spin Period, $P$ (s)   \dotfill & {0.118710251024231(5)}   \\  
Spin Period Derivative, $\dot{P}$ (s s$^{-1}$)                         \dotfill & {2.3056(3) $\times 10^{-18}$} \\  
Dispersion Measure, DM ($\mathrm{cm^{-3}\,pc}$)           \dotfill & {33.972(2)}              \\  
Orbital Period, $P_b$ (days)                          \dotfill & {95.19294198(4)}          \\  
Eccentricity, $e$                                         \dotfill & {0.00182570(2)}         \\  
Projected Semi-Major Axis, $x_{\rm p}$ (lt-s) \dotfill & {117.585776(1)}     \\  
Longitude of Periastron, $\omega$ ($^{\circ}$)                \dotfill & {99.5628(6)}         \\  
Epoch of Periastron (MJD)                          \dotfill & {60581.8944(2)}     \\  
Binary Model                                                  \dotfill & {DD}               \\
Solar System Ephemeris                                    \dotfill & {DE438}                \\
Terrestrial Time Standard                                 \dotfill & {TT(BIPM2023)}         \\
Time Units                                                \dotfill & {TDB}                  \\
\hline
Fitting parameters                                        & {}                    \\
First TOA (MJD)                                \dotfill & {60190.7895}           \\  
Last TOA (MJD)                                 \dotfill & {61090.3649}           \\  
Timing Epoch (MJD)                             \dotfill & {60558.8011}           \\  
Number of TOAs                                            \dotfill & {525}                  \\  
Weighted RMS Residual ($\mu\mathrm{s}$)                   \dotfill &{17.729}             \\
\hline
Derived parameters                                         & {}                    \\
Mass Function, $f(M_{\rm p})$ (${\rm M}_\odot$)           \dotfill &   0.192636 \\
Minimum companion mass, $M_{\rm c, min}$ (${\rm M}_\odot$)   \dotfill &   1.03   \\
Median companion mass, $M_{\rm c, med}$ (${\rm M}_\odot$)    &   1.27   \\
DM Distance, $d$ (${\rm kpc}$)                                    \dotfill & {1.76} ($\mathrm{NE2001}$) \\  
                                                                   & {1.96} ($\mathrm{YMW16}$)   \\
Characteristic Age, $\tau_c$ ($\mathrm{Myr}$)             \dotfill & {815.78}              \\  
Surface Magnetic Field, $B_{\mathrm{surf}}$ ($10^{10}\,\mathrm{G}$)  & {1.67}      \\  
Spin-Down Luminosity, $\dot{E}$ ($10^{30}\,\mathrm{erg\,s^{-1}}$)  \dotfill & {54.41}       \\
\hline
\label{table:timing_solution}
\end{tabular}  
\end{table}

\section{Results}\label{sec:3}

\subsection{Timing and improvement}\label{subsec:4}
\begin{figure*}
    \centering  
    \includegraphics[  
        width=1.0\textwidth]{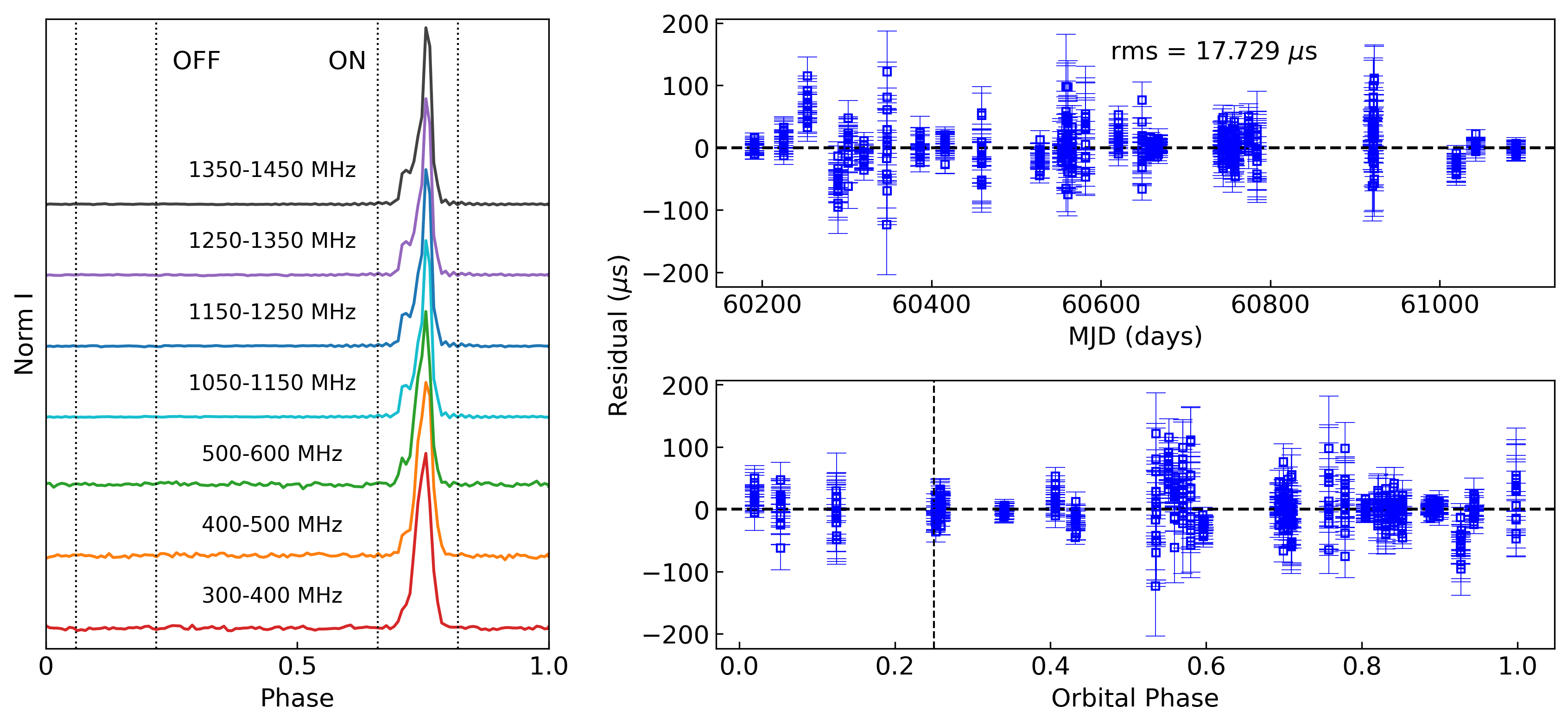}   
        \caption{Left: average pulse profiles at different frequencies. 
        Right: timing residuals of PSR~J2338+4818, including timing residuals as a function of MJD~(upper subplot) and orbital phase (lower subplot).
        A black dashed vertical line marks the orbital phase of 0.25.} 
    \label{fig:residual_profile}  
\end{figure*}

We utilized the \textsc{APTB}\footnote{\url{https://github.com/Camryn-Phillips/APT}} \citep{2024ApJ...964..128T}, based on \textsc{PINT}\footnote{\url{https://github.com/nanograv/PINT}} \citep{2021ApJ...911...45L,2024ApJ...971..150S},
to remove the phase jump in TOAs. 
Following the methodology of \textsc{APTB}, these jumps are arbitrary time offsets applied to independent observation clusters to account for the unknown phase counts (i.e., the global rotation counts) between them. 
By sequentially evaluating the phase wraps and removing these jumps, we successfully mapped the gaps between observations and obtained a unique phase-connected timing solution.
After obtaining a phase-connected timing solution using \textsc{APTB}, we incorporated TOAs from subsequent observational epochs into the existing dataset 
and performed a weighted least-squares fit with \textsc{PINT} to obtain the final timing parameters.
TOAs were weighted by their uncertainties, and the final timing solution has a reduced chi-square of 1.983.
Table~\ref{table:timing_solution} and Figure~\ref{fig:residual_profile} present the timing solution and the timing residuals, respectively.
Signals from PSR~J2338+4818 were detected in all observations, including those at orbital phase~0.25. 
No eclipse is detected.
The resulting calibrated average pulse profile is consistent with that reported by \citet{2021MNRAS.508..300C}.

The mass function is given by (e.g., \citealt{2016ARA&A..54..401O}):
\begin{equation}
    f(M_\mathrm{p},M_\mathrm{c}) = \frac{(M_\mathrm{c} \sin i)^3}{(M_\mathrm{p} + M_\mathrm{c})^2} = \frac{4\pi^2}{T_{\odot}}\frac{x_{\rm psr}^3}{P_\mathrm{b}^2}, 
    \label{eq:5}
\end{equation}
where $M_{\rm p}$ and $M_{\rm c}$ are the masses of pulsar and its companion, respectively, 
$i$ is the orbital inclination angle of the system, 
$T_{\odot} = G{M_{\odot}}/c^3 = 4.925490947\,\mu {\rm s}$ is the mass of the Sun in time units ($G$ is the Newton’s gravitational constant, $c$ is the speed of light and $M_{\odot}$ is the solar mass), $x_{\rm psr}$ is the projected semi-major axis and $P_{\rm b}$ is the orbital period.
Assuming $M_\mathrm{p}\, = \, 1.35 \, M_{\odot}$, 
we obtain a minimum ($i\, = \,90^\circ$) and median ($i\, = \,60^\circ$) companion masses of 1.03 and 1.27~$M_{\odot}$, respectively.

The estimated companion mass of PSR~J2338+4818 from timing lies between 1.03 and 1.27~$\rm M_{\odot}$.
We have checked the optical image at the PSR~J2338+4818 position but find no counterpart. 
This suggests that the companion of PSR~J2338+4818 is likely a compact star.
PSR~J2338+4818 is in a wide binary in a nearly circular orbit with a period of roughly 95.19~days, suggesting that the companion is more likely a white dwarf rather than a neutron star, since the supernova explosion to generate a neutron star would destroy the wide binary \citep{1991PhR...203....1B,2017ApJ...846..170T}.
The estimated minimum companion mass is consistent with a CO white dwarf, which has the mass range of 0.5 to 1.1~$M_{\odot}$ \citep{1990ASPC...11..483I}.
On the other hand, the characteristic age of PSR~J2338+4818 is $\sim 0.82 {\ \rm Gyr}$ (estimated according to \cite{2004hpa..book.....L}).
This timescale is slightly longer than the time required for a main-sequence star of 3 to 6~$M_{\odot}$ to evolve to a CO white dwarf, 
which is about 0.3 to 0.6~{\rm Gyr} \citep[see][]{2019MNRAS.490.2013C}. 

\begin{figure}
    \centering  
    \includegraphics[width=0.75\textwidth, clip, trim=10 10 10 100]{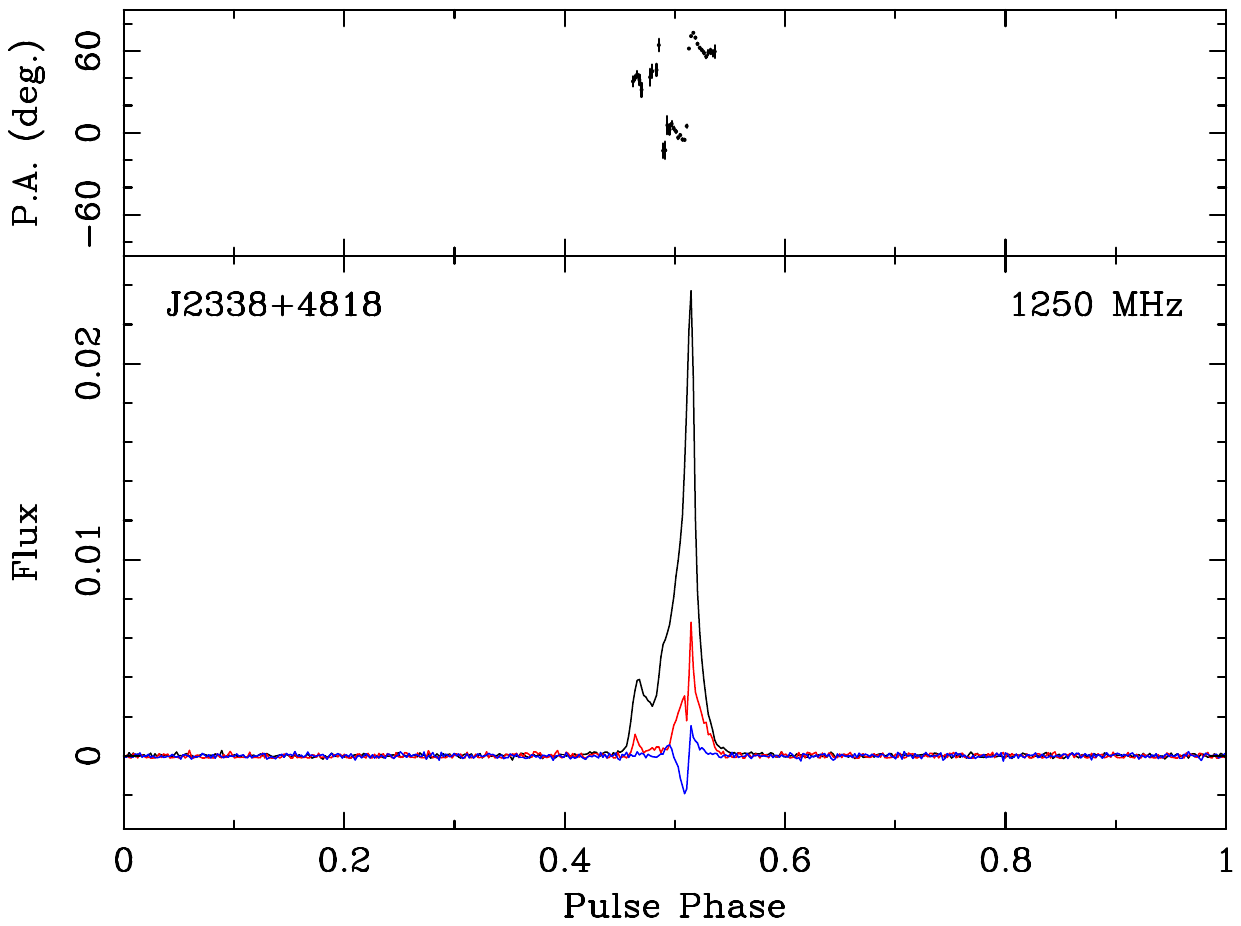}  
    \caption{The calibrated polarisation profile of PSR~J2338+4818 at the central frequency of 1250~{\rm MHz} from the observation on MJD~60553. The total intensity ($I$) is shown in black, the linear polarisation ($L$) in red, and the circular polarisation ($V$) in blue. The polarisation position angle is shown in the top panel.}
    \label{fig:polarisation_profile}  
\end{figure}

We obtained the ${RM} = -16 \pm 1 \, {\rm rad\, m^{-2}}$ on the observation in MJD~60553, 
and the calibrated polarisation profile is presented in Figure~\ref{fig:polarisation_profile}.

\subsection{Single-pulse and nulling}\label{subsec:5}
\begin{figure}
    \centering  
    \includegraphics[width=\textwidth]{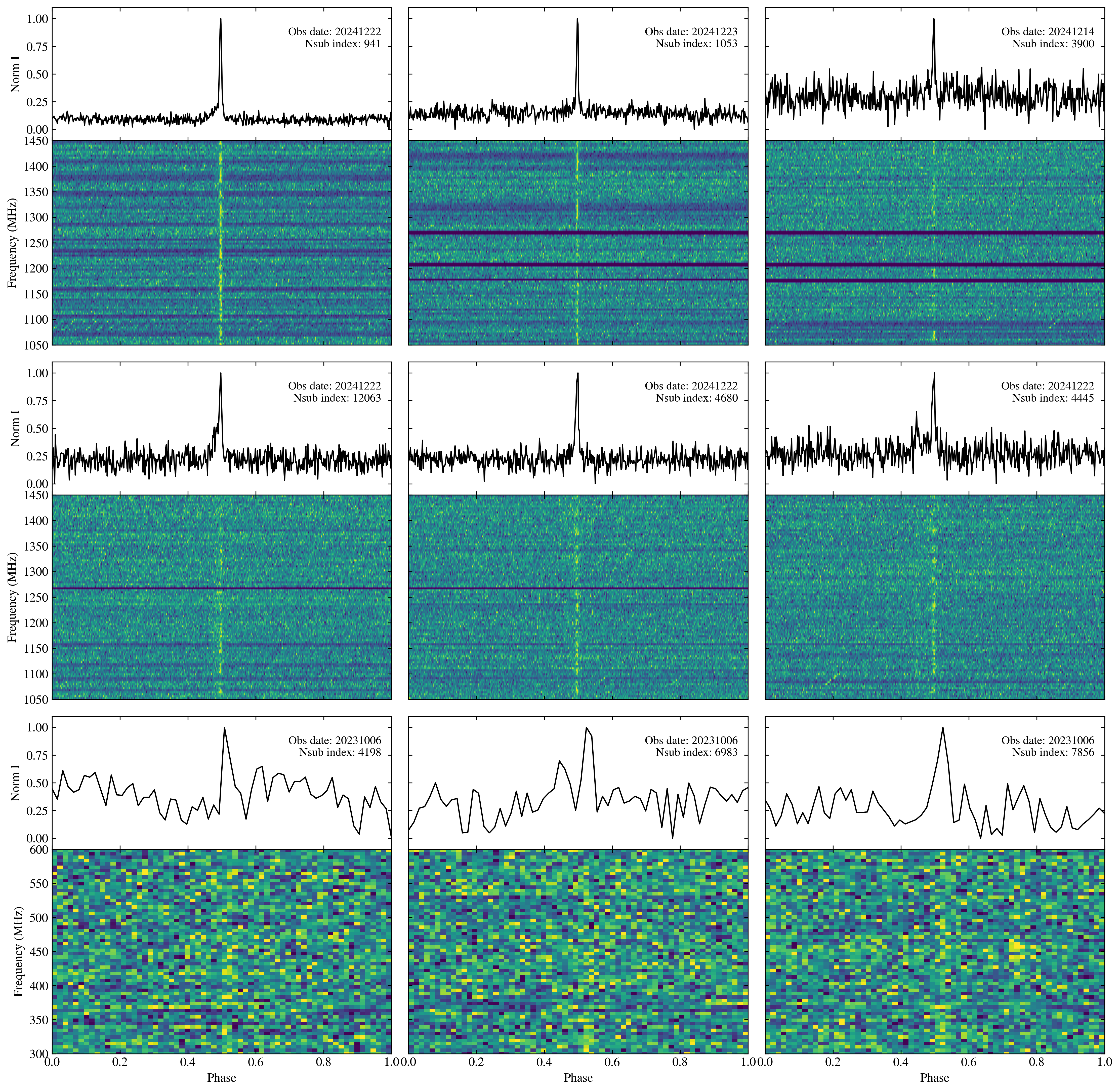}  
    \caption{The 9~single pulses from PSR~J2338+4818. Each subpanel consists of a normalized pulse profile and a frequency-phase diagram. The first row, from left to right, shows the three single pulses with the highest S/N from MJD~60666, MJD~60667, and MJD~60658. The second row displays three single pulses from MJD~60666, where scintillation is clearly noticeable at different times. The third row presents three single pulses from low-frequency data from MJD~60222.} 
    \label{fig:singlepulse}  
\end{figure}

We analyse the single-pulse properties using our observations. 
The distributions of single-pulse amplitudes and ${\rm S/N}$ are used to investigate emission phenomena such as giant pulses, 
which are typically defined as pulses with flux densities exceeding ten times the mean flux density \citep{2012A&A...538A...7K}. 
For each single pulse, the ${\rm S/N}$ is calculated as \citep{2023MNRAS.520.2747P}
\begin{equation}
\mathrm{S/N} = \frac{I_{\rm peak}}{\sigma_p},
\end{equation}
where $I_{\rm peak}$ is the peak intensity of the pulse and $\sigma_p$ is the standard deviation of the off-pulse region, representing the noise level of the observation. 
In total, we have detected 27,228 single pulses with $\mathrm{S/N} > 7$ from PSR~J2338+4818 in 36~observations.
The total number and fraction of such detected single pulses are listed in Table~\ref{table:FAST_observations}.
In Figure~\ref{fig:singlepulse}, we show 9~single pulses from PSR~J2338+4818.
The first row displays 3 single pulses with the highest S/N from different observations. 
The second row presents 3 single pulses from MJD~60666, where scintillation is clearly noticeable. 
The third row shows 3 single pulses from MJD~60222.
No giant pulses exceeding 10~times the mean peak amplitude are found in our observations.

To investigate possible pulse nulling in PSR~J2338+4818, we analyzed the single-pulse energy distributions. 
The ON-pulse and OFF-pulse windows were determined from the average pulse profile after removing the baseline. 
The ON-pulse window encompasses the full phase range of detectable emission, 
while the OFF-pulse window has the same phase width but is selected from a region free of pulse emission (see the left panel of Figure~\ref{fig:residual_profile} for the ON-pulse and OFF-pulse windows).
For each pulse, the ON-pulse energy was computed by summing the intensities within the ON-pulse window after subtracting the baseline level estimated from the mean OFF-pulse intensity. 
The OFF-pulse energy was calculated in the same manner using an equal number of bins from the OFF-pulse region. 
All energies were normalized by the mean ON-pulse energy, yielding the relative pulse energy.

Histograms of the ON-pulse and OFF-pulse energies were constructed to characterize the emission and noise distributions. 
A classical estimate of the nulling fraction (NF) was obtained using the subtraction method of \citet{1976MNRAS.176..249R}, in which a scaled OFF-pulse histogram is subtracted from the ON-pulse histogram. 
However, as demonstrated by \citet{2018ApJ...855...14K}, this method can produce biased NF estimates when the emission component is comparable to the noise level.
We adopted the probabilistic mixture-model approach described by \citet{2023ApJ...948...32A}. 
The single-pulse energies $x$ are treated as random draws from a probability density function
\begin{equation}
p(x \mid \boldsymbol{\theta}) = \sum_{n=1}^{m} c_n F_n(x \mid \boldsymbol{\theta}_n),
\end{equation}
where $F_n$ are the component probability density functions parameterized by $\boldsymbol{\theta}_n$, $c_n$ are their weights, and $\sum_{n=1}^{m} c_n = 1$. 
In the Gaussian case,
\begin{equation}
F_n(x;\mu_n,\sigma_n) = \frac{1}{\sqrt{2\pi}\sigma_n}
\exp\left[-\frac{(x-\mu_n)^2}{2\sigma_n^2}\right],
\end{equation}
the model reduces to a Gaussian mixture model (GMM).

The OFF-pulse energy distribution is well described by a Gaussian, consistent with radiometer noise, and was used to inform the parameters of the null component. 
The emission component can deviate from a pure Gaussian due to interstellar scattering, which introduces exponential tails \citep{2023ApJ...948...32A}
\begin{equation}
F(x;\mu,\sigma,\tau) = \frac{1}{2\tau} \exp\left(\frac{\sigma^2}{2\tau^2}
- \frac{x-\mu}{\tau}\right)
\operatorname{erfc}\!\left(\frac{x-(\mu+\sigma^2/\tau)}{\sqrt{2}\sigma}\right),
\end{equation}
where $\tau$ is the exponential decay time. 
Different models were ranked using the Bayesian Information Criterion (BIC), and the best-fitting model was selected. 
The parameters were estimated using Markov Chain Monte Carlo (MCMC; e.g., \citealt{2013PASP..125..306F}) sampling following \citet{2018ApJ...855...14K} and \citet{2023ApJ...948...32A}.

\begin{figure}
    \centering  
    \includegraphics[width=0.65\textwidth]{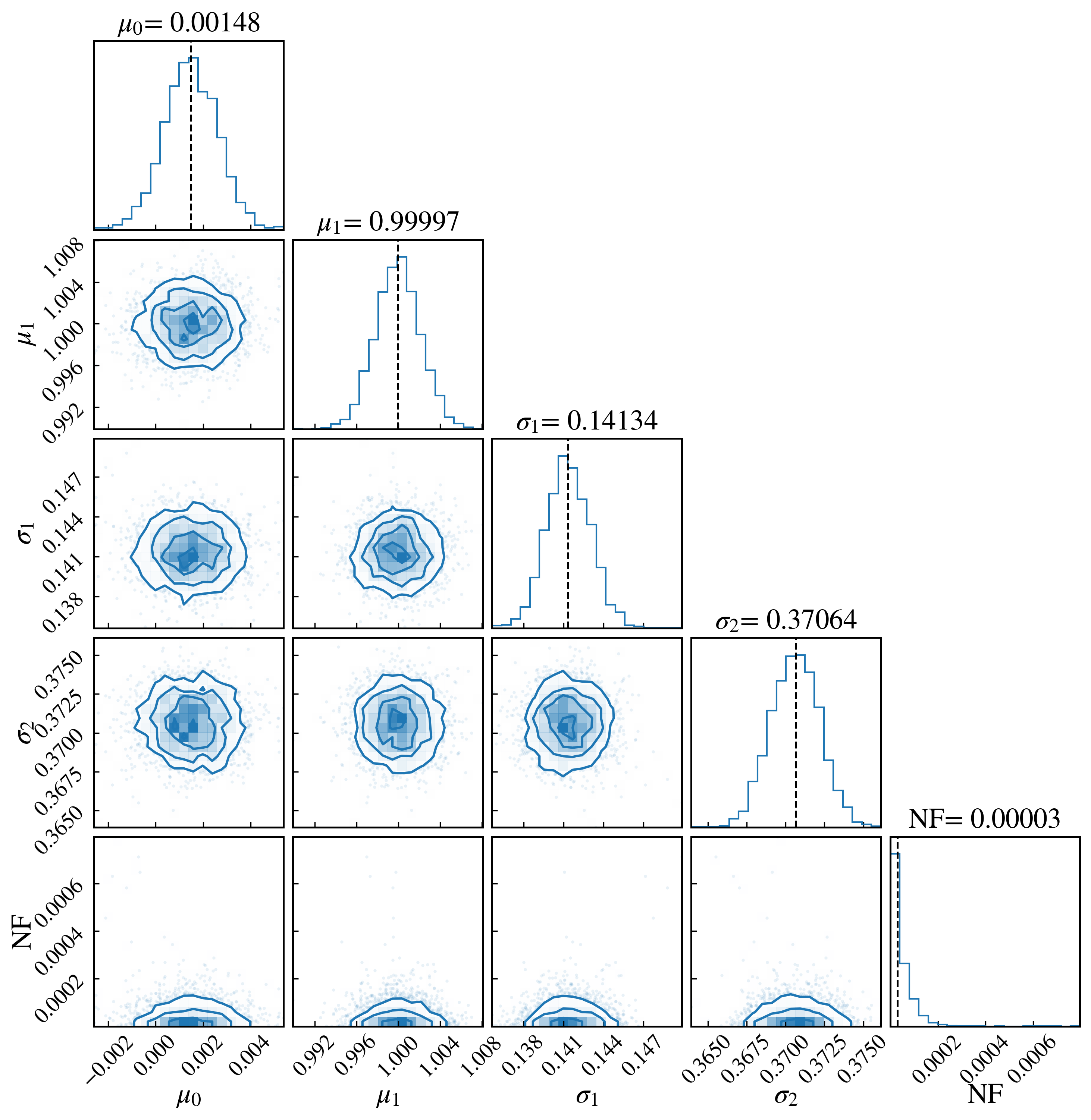}  
        \caption{Posterior probability distributions (derived from the observation at MJD~60666) of PSR~J2338+4818 of the GMM parameters for the null and emission components, derived using the MCMC algorithm (see \citealt{2018ApJ...855...14K}; \citealt{2023ApJ...948...32A}). The diagonal panels show the marginalized one-dimensional posteriors for each parameter, with vertical dashed lines marking the median values.} 
    \label{fig:MCMC}  
\end{figure}

Figure~\ref{fig:MCMC} shows the posterior probability distributions of the GMM parameters derived from the observation at MJD~60666. 
We obtain ${\rm NF} < 0.003~\%$ for the observation at MJD~60666.
The null component contributes negligibly to the total distribution, indicating that the nulling fraction is consistent with zero within uncertainties. 

\begin{figure}
    \centering  
    \includegraphics[width=0.5\textwidth]{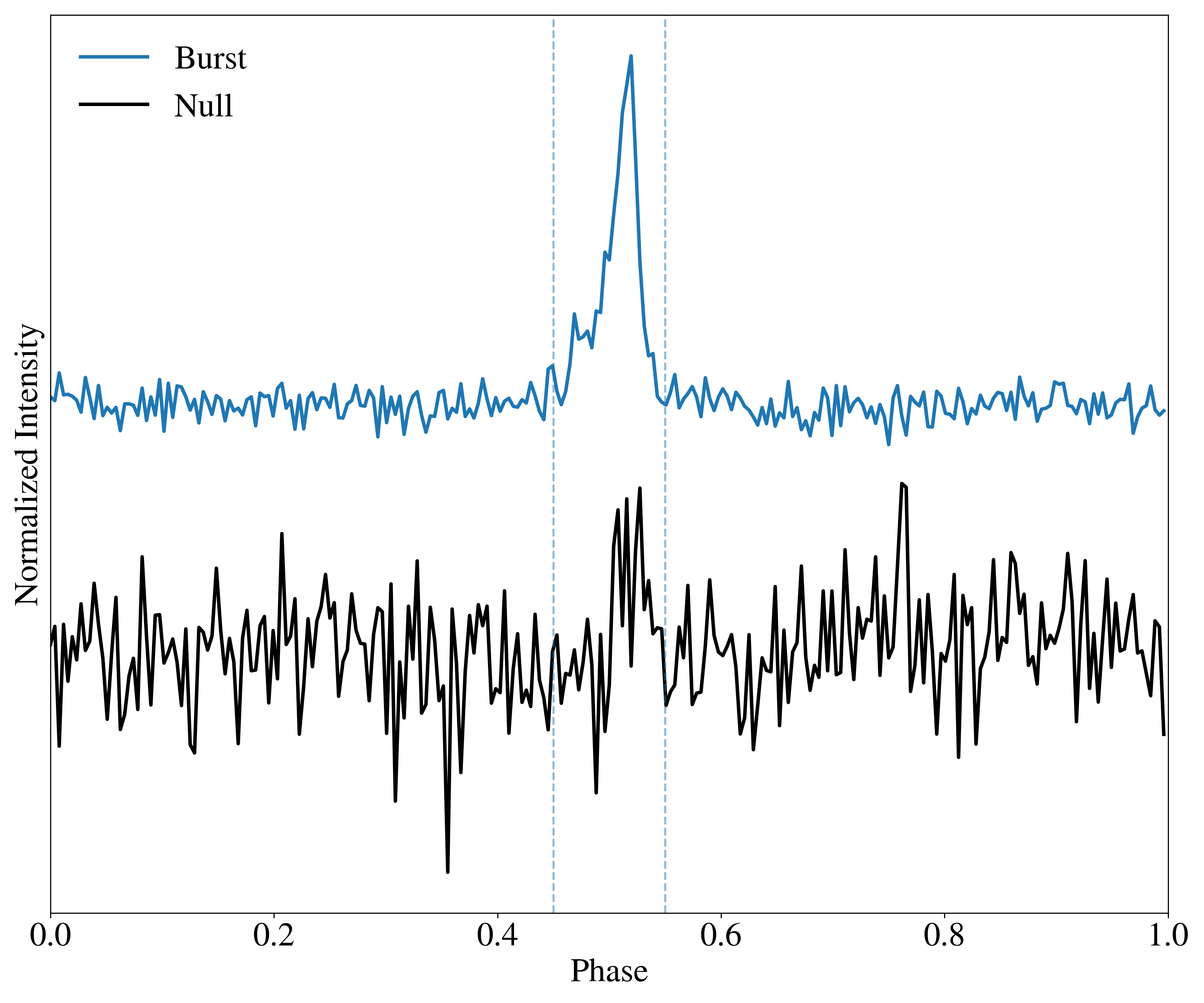} 
    \caption{Comparison of the integrated profiles for pulses identified as bursts ($P_{\rm null} < 0.5$; N $=59,569$) and nulls ($P_{\rm null} \ge 0.5$; N $= 234$) for the observation on MJD~60347, where the profiles consist of 256 phase bins and are normalized for comparison.} 
    \label{fig:C1_null_burst_comparison}  
\end{figure}

Across the all observations except the ultra-wideband observation, 30~observations have an NF less than 10~\%, four observations have an NF between 10~\% and 20~\%, and one observation has an NF greater than 20~\% (see Table~\ref{table:FAST_observations}).
In low ${\rm S/N}$ sessions (e.g., MJD~60347, where NF $\approx 36~\%$), the single-pulse energy distribution significantly overlaps with the noise background, and no distinct null component is resolved in the pulse stacks. 
We therefore attribute these elevated NF estimates to statistical fluctuations and model degeneracy in low ${\rm S/N}$ data \citep{2018ApJ...855...14K}, rather than to intrinsic pulse nulling.
To verify this interpretation, we performed a stacking analysis of pulses classified as nulls and bursts and examined the corresponding integrated profiles.  
The classification was based on the null probabilities of individual pulses ($P_{\rm null}$), with pulses having $P_{\rm null} \ge 0.5$ assigned to the null component and those with $P_{\rm null} < 0.5$ assigned to the burst component.  
For the MJD~60347 observation, 2,595 pulses were classified as nulls, while the remaining 57,208 pulses were classified as bursts.  
In Figure~\ref{fig:C1_null_burst_comparison}, we show the average profiles of the null and burst components obtained from the observation at MJD~60347.
While the burst component exhibits a highly significant pulse profile, the averaged null component shows a very weak feature at the expected pulse phase.
To examine whether this feature corresponds to genuine weak emission or random noise, we investigated the dependence of the $\rm S/N$ of the stacked null-component pulses on the integration time.
The null pulses were randomly divided into subsets corresponding to one-quarter, one-half, and the full sample, and stacked separately to form integrated profiles for each subset, with each individual pulse having an identical integration time.
The S/N in the ON-pulse window for stacked subsets of null pulses was found to follow an approximate $\mathrm{S/N} \propto \sqrt{N}$ relation, where N is the number of integrated null pulses.
To further verify this result, we performed an identical stacking analysis on an OFF-pulse phase window. 
The measured $\rm S/N$ remained consistent with zero for all integration lengths and showed no systematic dependence on $\sqrt{N}$.
These results provide evidence that the null component contains a weak emission component at the ON-pulse phase.

\begin{figure}
    \centering  
    \includegraphics[width=\textwidth]{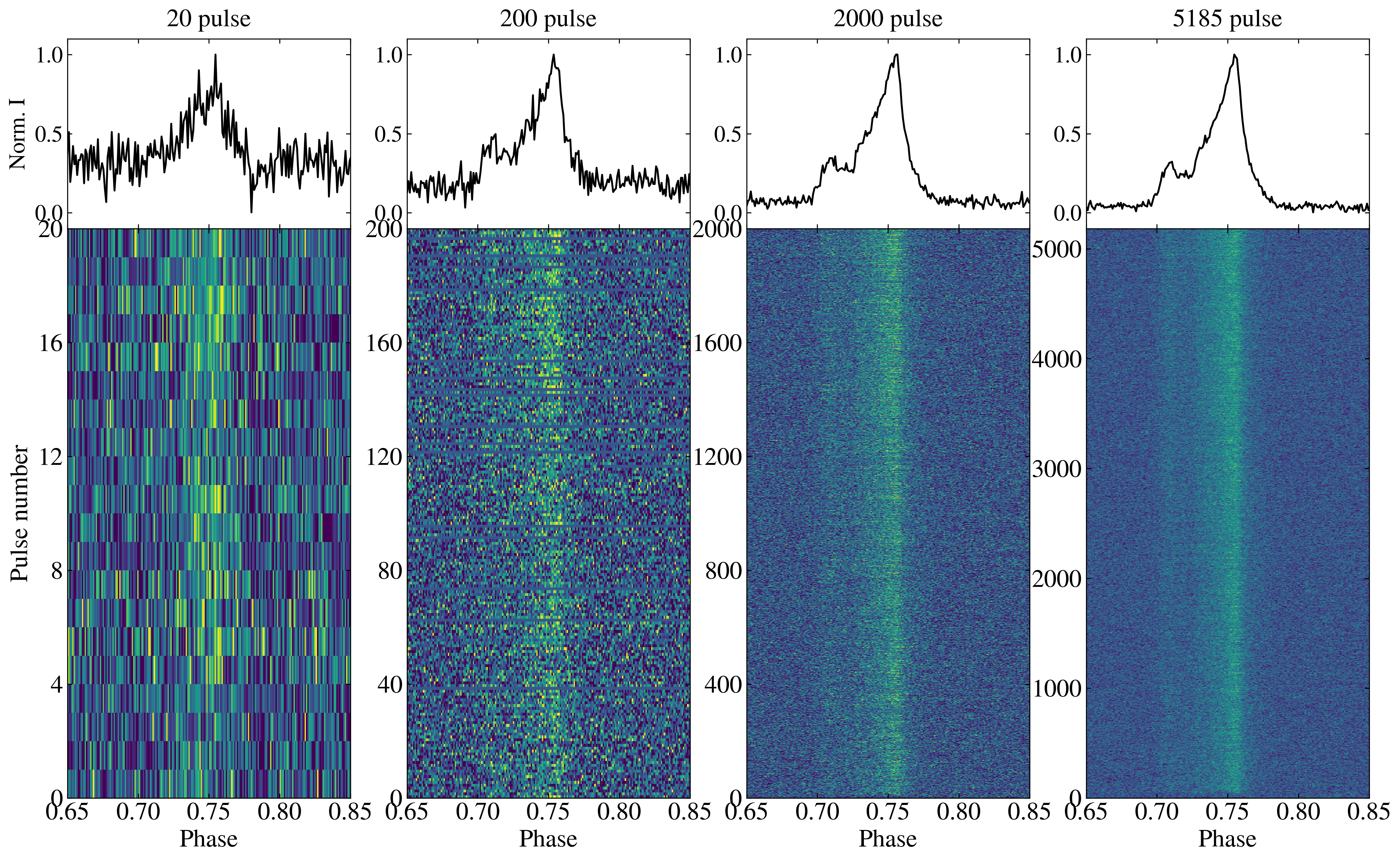} 
    \caption{Normalized average pulse profiles (top) and stacked single pulses (bottom) for low ${\rm S/N}$ pulses from the observation at MJD~60743. From left to right, the number of stacked pulses is 20, 200, 2000, and the total number of pulses satisfying $\mathrm{S/N} < 3$.} 
    \label{fig:C1_diff_singlepulse_stack}  
\end{figure} 

Furthermore, for each observation, we selected single pulses with $\mathrm{S/N} < 5$ and $\mathrm{S/N} < 3$, respectively, and stacked different numbers of these pulses to obtain the corresponding normalized average pulse profiles.
As shown in Figure~\ref{fig:C1_diff_singlepulse_stack}, we present an example from the observation at MJD 60743, where single pulses with $\mathrm{S/N} < 3$ were selected and stacked using 20, 200, 2000 pulses, and the total number of pulses with $\mathrm{S/N} < 3$, to produce the corresponding normalized average pulse profiles.
For this observation, out of a total of 15,296 single pulses, 5,185 pulses have $\mathrm{S/N} < 3$, 
corresponding to 33.72~\% of the total.
Even a small number of low ${\rm S/N}$ pulses can produce a clearly visible pulse profile.
As the number of stacked pulses increases, the average pulse profile becomes clearer and more stable.
This further confirms that even at low ${\rm S/N}$, the ON-pulse region is predominantly composed of emission, while the null component is almost negligible.

\begin{figure}
    \centering  
    \includegraphics[width=0.65\textwidth]{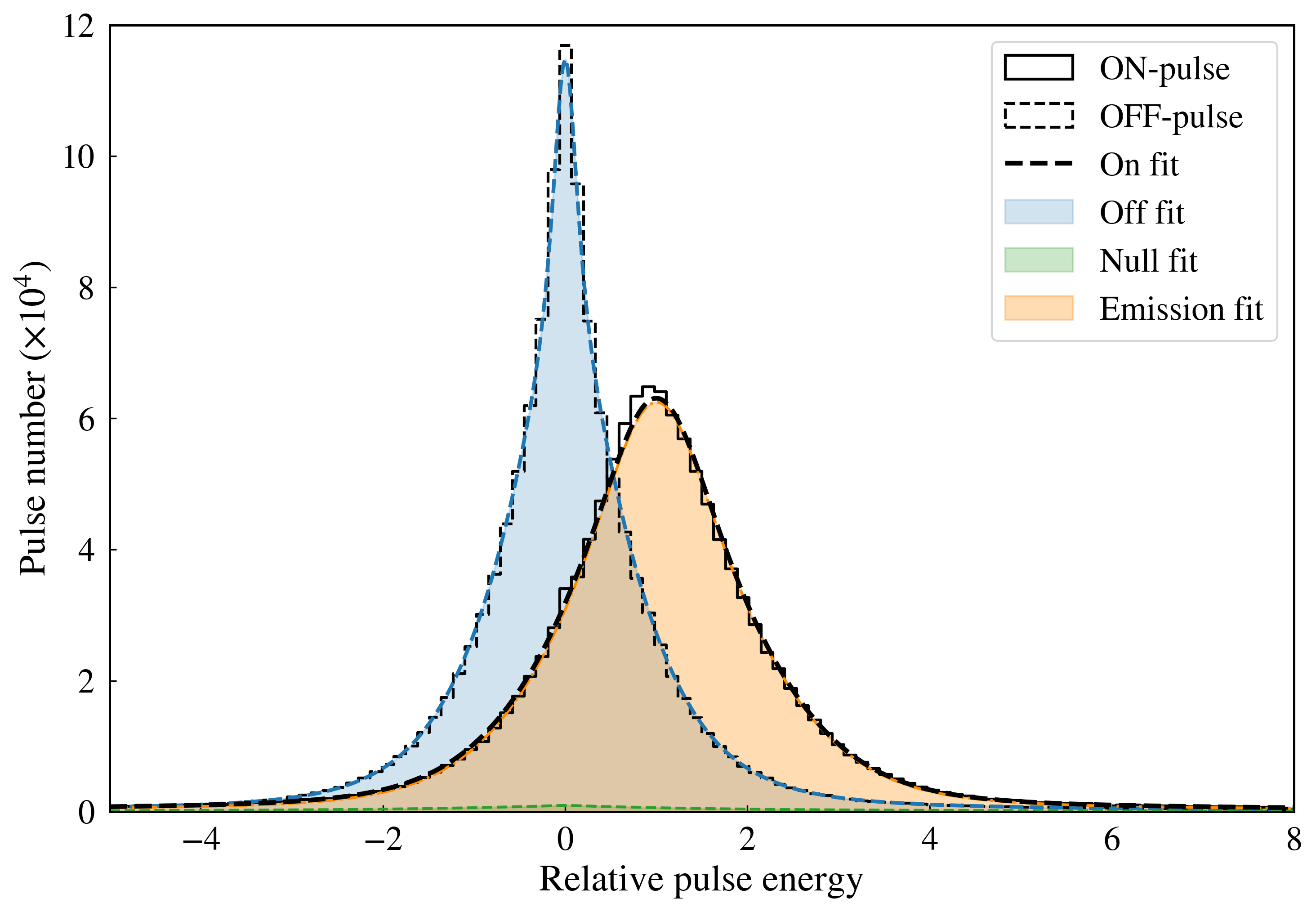} 
    \caption{Pulse energy histograms (derived from the 35 observations except the ultra-wideband observation) for the ON-pulse and OFF-pulse windows of PSR~J2338+4818. Individual ON and OFF histograms are shown as black hollow and dashed black boxes, respectively. The best-fitting GMM components are shown as filled regions: the null component (blue), the emission component (orange), and the sum of the null and emission components (black dotted line).} 
    \label{fig:GMM}  
\end{figure}

Figure~\ref{fig:GMM} presents the pulse energy histograms and best-fitting mixture models for all observations. 
The emission component clearly dominates the distribution, while the null component is nearly negligible, confirming that PSR~J2338+4818 does not exhibit significant pulse nulling.

By examining the pulse energy histograms from 35 observations except the ultra-wideband observation, we find that the null component is nearly absent, with the majority of the distribution being dominated by the emission component. 
This further supports the conclusion that we cannot confirm the long-term nulling phenomenon suggested by \cite{2021MNRAS.508..300C}. 
PSR~J2338+4818 is unlikely to exhibit the significant nulling behavior in FAST observations.

We present single-pulse data from PSR~J2338+4818 across different observations and time periods in Figure~\ref{fig:singlepulse}. 
It shows the impact of scintillation on the pulse signal, and the observed variations are more likely due to interstellar scintillation rather than intrinsic nulling.
The long-term nulling phenomenon proposed by \cite{2021MNRAS.508..300C}, 
may be a result of strong interstellar scintillation, and further observations are needed to confirm it.

\begin{table}
\centering 
\caption{The Scintillation Parameters of PSR~J2338+4818}
\begin{tabular*}{\textwidth}{@{\extracolsep{\fill}}cccccccccccc@{}}
\hline
MJD & \(A\) & \(W\) & \(\Delta {\tau }_{\mathrm{d}}\) (min) & \(\Delta {\nu }_{\mathrm{d}}\) (${\rm MHz}$) & B (${\rm MHz}$) & $f_c$ (${\rm MHz}$) & \({\sigma }_{r}\) & \(u\) \\
\hline\hline
60190 & ${0.787} \pm {0.002}$ & ${0.213} \pm {0.002}$ & ${12.97} \pm {0.06}$ & ${9.29} \pm {0.13}$ & 400 & 1250 & 0.14 & 16 \\
60253 & ${0.687} \pm {0.004}$ & ${0.313} \pm {0.004}$ & ${15.35} \pm {0.14}$ & ${14.04} \pm {0.24}$ & 400 & 1250 & 0.19 & 13 \\
60289 & ${0.572} \pm {0.002}$ & ${0.428} \pm {0.002}$ & ${17.18} \pm {0.08}$ & ${2.11} \pm {0.06}$ & 300 & 1250 & 0.06 & 34 \\
60301 & ${0.735} \pm {0.004}$ & ${0.265} \pm {0.004}$ & ${14.25} \pm {0.13}$ & ${7.61} \pm {0.10}$ & 400 & 1250 & 0.14 & 18 \\
60320 & ${0.578} \pm {0.002}$ & ${0.422} \pm {0.002}$ & ${8.23} \pm {0.04}$ & ${2.02} \pm {0.05}$ & 400 & 1250 & 0.04 & 35 \\
60527 & ${0.589} \pm {0.001}$ & ${0.411} \pm {0.001}$ & ${13.32} \pm {0.05}$ & ${27.41} \pm {0.52}$ & 400 & 1250 & 0.17 & 10 \\
60553 & ${0.707} \pm {0.007}$ & ${0.293} \pm {0.007}$ & ${2.93} \pm {0.05}$ & ${4.89} \pm {0.11}$ & 400 & 1250 & 0.08 & 23 \\
60620 & ${0.866} \pm {0.004}$ & ${0.134} \pm {0.004}$ & ${7.58} \pm {0.05}$ & ${2.72} \pm {0.04}$ & 400 & 1250 & 0.08 & 30 \\
60658 & ${0.937} \pm {0.001}$ & ${0.063} \pm {0.001}$ & ${12.93} \pm {0.02}$ & ${3.18} \pm {0.05}$ & 400 & 1250 & 0.08 & 28 \\
60666 & ${0.981} \pm {0.001}$ & ${0.019} \pm {0.001}$ & ${13.17} \pm {0.02}$ & ${1.68} \pm {0.03}$ & 400 & 1250 & 0.06 & 39 \\
60667 & ${0.932} \pm {0.001}$ & ${0.068} \pm {0.001}$ & ${12.30} \pm {0.03}$ & ${3.22} \pm {0.05}$ & 400 & 1250 & 0.08 & 28 \\
60755 & ${1.151} \pm {0.002}$ & ${-0.151} \pm {0.002}$ & ${25.26} \pm {0.08}$ & ${18.93} \pm {0.28}$ & 130 & 1435 & 0.36 & 12 \\
\hline\hline
\end{tabular*}
\label{table:scintillation_parameters}
\begin{flushleft} 
Note: $A$ is a constant fitted in Equation~\ref{eq:8}, and $W$ is the white noise spike. $\Delta {\tau }_{\rm d}$ and $\Delta {\nu }_{\rm d}$ are the scintillation timescale and bandwidth, respectively. ${\sigma }_{r}$ is the uncertainty in the fitted parameters. $B$ is the effective bandwidth after removing edge channels and RFI. $u$ is the scintillation strength. 
In the observation on MJD~60289, RFI was present in the 100~${\rm MHz}$ bandwidth at both the upper and lower edges, reducing the effective bandwidth to 300~${\rm MHz}$. Severe RFI at 1250~${\rm MHz}$ on MJD~60775 prompted an adjustment of the center frequency to 1435~MHz, resulting in a narrower bandwidth of 130~${\rm MHz}$, over which the auto-correlation function was well fitted.
\end{flushleft}
\end{table}

\subsection{Scintillation}\label{subsec:6}
The inhomogeneity of electron density in the interstellar medium leads to the scattering of radio signals radiated and produces diffraction patterns. 
The relative motions between pulsars, the Earth, and the interstellar medium cause diffraction patterns to sweep across the Earth, resulting in the change of flux density at different time scales, which is the scintillation phenomenon of pulsars.

The scintillation bandwidth $\Delta {\nu }_{\mathrm{d}}$ and the scintillation timescale $\Delta {\tau }_{\mathrm{d}}$ can be used to quantify the average scintillation characteristics of each observation. 
We obtained $\Delta {\tau }_{\mathrm{d}}$ by using a least-square fit to the one-dimensional time-domain ACF \citep{2019MNRAS.485.4389R},
\begin{equation}
C\left( {{\Delta \tau },0}\right)  = C\left( {0,0}\right)\exp \left( {-{\left\vert \frac{\Delta \tau }{\Delta {\tau }_{\rm d}}\right\vert }^{\frac{5}{3}}}\right) ,
\label{eq:7}
\end{equation}
and
\begin{equation}
C\left( {0,0}\right)  = A + W, 
\label{eq:8}
\end{equation}
where $A$ is a constant to be fitted and $W$ is the white noise spike. 
After obtaining $A$ and $W$ , we fixed $A$ and got $\Delta {\nu }_{\mathrm{d}}$ by doing a least-squares fit to the one-dimensional frequency-domain ACF via,
\begin{equation}
C\left( {0,{\Delta \nu }}\right)  = C\left( {0,0}\right)\exp \left( {-\left| \frac{\Delta \nu }{\Delta {\nu }_{\mathrm{d}}/\ln 2}\right| }\right),  
\end{equation}
The scintillation parameters $\nu_d$ and $\tau_d$ are obtained from the fitted Gaussian model parameters $C_1$ and $C_3$, respectively, as follows \citep{1999ApJS..121..483B}:
\begin{equation}
\nu_d = \left( \frac{\ln 2}{C_1} \right)^{0.5}, \quad \tau_d = \left( \frac{1}{C_3} \right)^{0.5}, 
\end{equation}
These decorrelation widths $\nu_d$ and $\tau_d$ are corrected for smearing effects due to finite instrumental resolutions in frequency and time, using a quadrature subtraction scheme. 
In terms of the fitted Gaussian parameters, the slope $dt/d\nu$, which characterizes the drift in the dynamic spectrum, can be expressed as \citep{1999ApJS..121..483B}:
\begin{equation}
\frac{dt}{d\nu} = - \frac{C_2}{2C_3}.
\end{equation}
The uncertainties in $C_1$, $C_2$, and $C_3$ due to the Gaussian model fitting are obtained through $\chi^2$ analysis and are referred to as $\sigma_{\mathrm{mod}}$.
These uncertainties are propagated into the final uncertainties of the scintillation parameters, including $\nu_d$, $\tau_d$, and $dt/d\nu$.
For the parameters $\nu_d$, $\tau_d$, and $dt/d\nu$, the errors from the Gaussian fitting ($\sigma_{\mathrm{mod}}$) are added in quadrature with the statistical errors, $\sigma_{\mathrm{r}}$, arising from the finite number of scintles, to obtain their final uncertainties. 
The statistical uncertainty is given by \citep{1999ApJS..121..483B}:
\begin{equation}
{\sigma }_{r} = {\left( \frac{fBT}{\Delta {\nu }_{\mathrm{d}}\Delta {\tau }_{\mathrm{d}}}\right) }^{-{0.5}}, 
\end{equation}
where $T$ and $B$ are the integration time and bandwidth, respectively. The filling factor $f$ is the percentage of scintillation area in the dynamic spectrum. We calculated the percentage of data with more than 1~$\sigma$ in the dynamic spectra and chose $f$ = 0.25 in the study.
From \citet{1990MNRAS.244...68R} and \citet{1999ApJS..121..483B}, the scintillation strength is
\begin{equation}
u \approx  \sqrt{\frac{2\nu }{\Delta {\nu }_{\mathrm{d}}}},
\end{equation}
where $\nu$ is the central frequency (1250~${\rm MHz}$). When $u < 1$, it shows weak scintillation. When $u > 1$, it indicates strong scintillation.

We could also search for arcs in the secondary spectrum. The scintillation arc can be described as in \citet{2019MNRAS.485.4389R}:
\begin{equation}
{f}_{\lambda } = \eta {f}_{t}^{2}, 
\end{equation}
in which
\begin{equation}
\eta  = {1.543} \times {10}^{7}\frac{{D}_{\mathrm{{kpc}}}s\left( {1 - s}\right) }{{\left( {V}_{\mathrm{{eff}}, \bot}\cos \psi \right) }^{2}}, 
\label{eq:13}
\end{equation}
where $\eta$ is the arc curvature, ${D}_{\mathrm{{kpc}}}$ is the pulsar distance in $\mathrm{{kpc}}$, $s$ is the fractional screen distance \citep[see e.g.,][]{2025NatAs...9.1053R}, $\psi$ is the angle between the major axis of the anisotropic structure and the effective velocity vector, and ${V}_{\text{eff,} \bot}$ is the effective perpendicular velocity in kilometers per second ($\mathrm{km\,s^{-1}}$). In all observations, the secondary spectra of PSR~J2338+4818 show no evidence of a clear scintillation arc.

\begin{figure*}
    \centering  
    \includegraphics[  
        width=\textwidth]{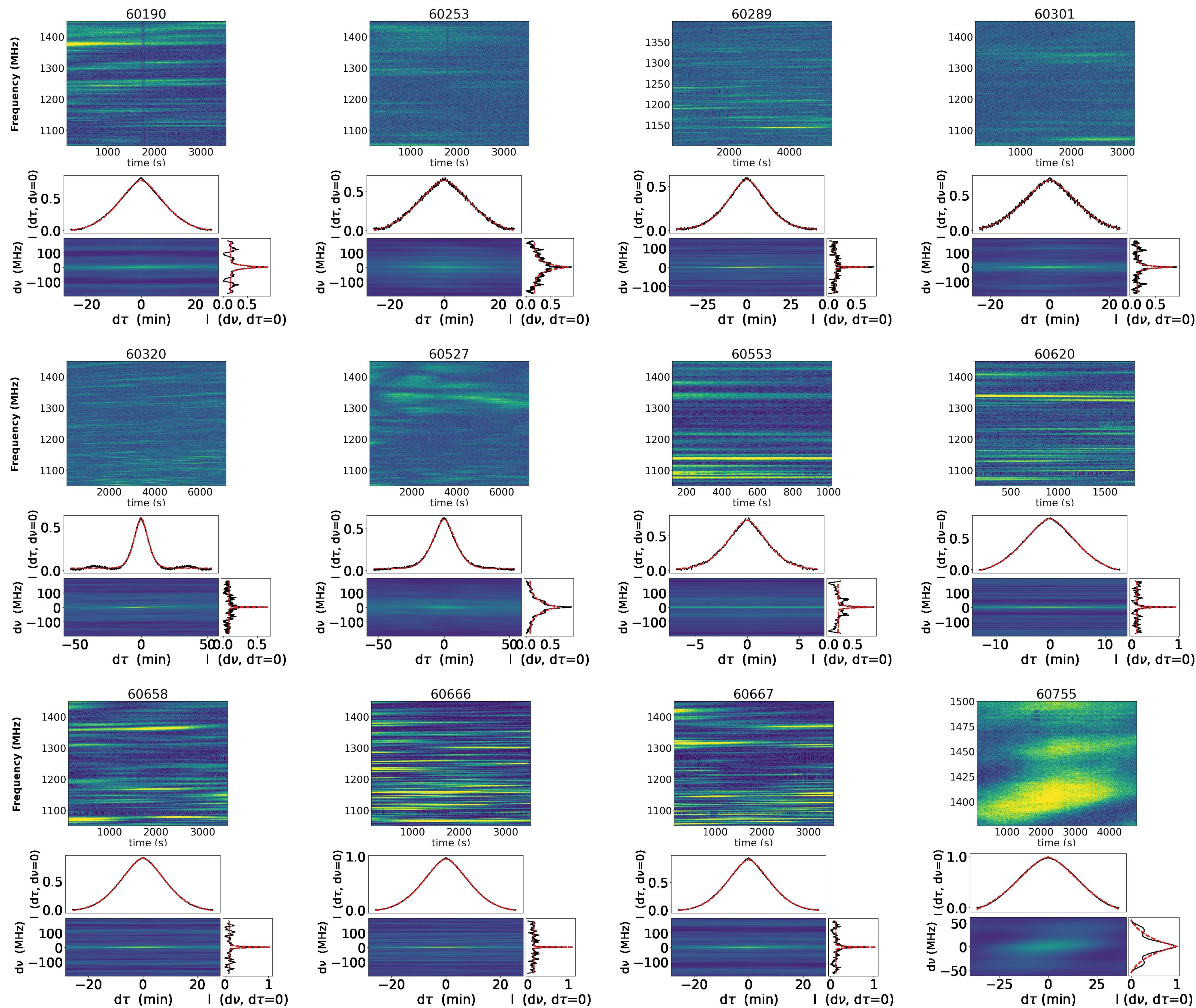}  
        \caption{Dynamic spectra (top panels), ACFs (bottom panels) of PSR~J2338+4818. The integration times for these observations range from 1200 to 7800~${\rm s}$. In these one-dimensional time-domain and frequency-domain ACFs, the red lines show the best-fit results for \(\Delta \tau_d\) and \(\Delta \nu_d\), respectively.} 
    \label{fig:ACF_dynamic_spectra}  
\end{figure*}

We performed ACF analysis on all observations and obtained the scintillation parameters for PSR~J2338+4818 from those observations where the parameters could be reliably fitted.
The results are presented in Table~\ref{table:scintillation_parameters}.
In Figure~\ref{fig:ACF_dynamic_spectra},
we showed the dynamic spectra and ACFs for PSR~J2338+4818. 
It is noted that the limited number of patches captured in most dynamic spectra may introduce additional errors in the measurement of these parameters.

From Table~\ref{table:scintillation_parameters}, 
we found that the scintillation bandwidth $\Delta \nu_d$ exhibits substantial variation across epochs, 
ranging from 1.68~${\rm MHz}$ (MJD~60666) to 27.41~${\rm MHz}$ (MJD~60527). 
This variation indicates significant inhomogeneities in the interstellar medium (ISM) along the line of sight to the pulsar.
The scintillation timescale $\Delta \tau_d$ ranges from 2.93~minutes (MJD~60553) to 25.26~minutes (MJD~60755), with the majority of observations falling within the range of several to tens of minutes. 
This suggests that the ISM along the line of sight to the pulsar may be dominated by larger and more stable scattering regions.

For pulsars in binary systems, multiple observations at different orbital phases can help constrain orbital parameters by studying the scintillation timescale and the variation of arc curvature across orbital phases \citep{2019MNRAS.485.4389R}.
None of our observations show clear arc curvature.
With future observations, we may be able to detect the arc curvature and determine the location of the scattering screen of PSR~J2338+4818.

\section{Discussion and Conclusions}\label{Conclusions}

\subsection{Discussion}

\subsubsection{Rate of Periastron Advance}

\begin{figure}
    \centering  
    \includegraphics[width=0.75\textwidth]{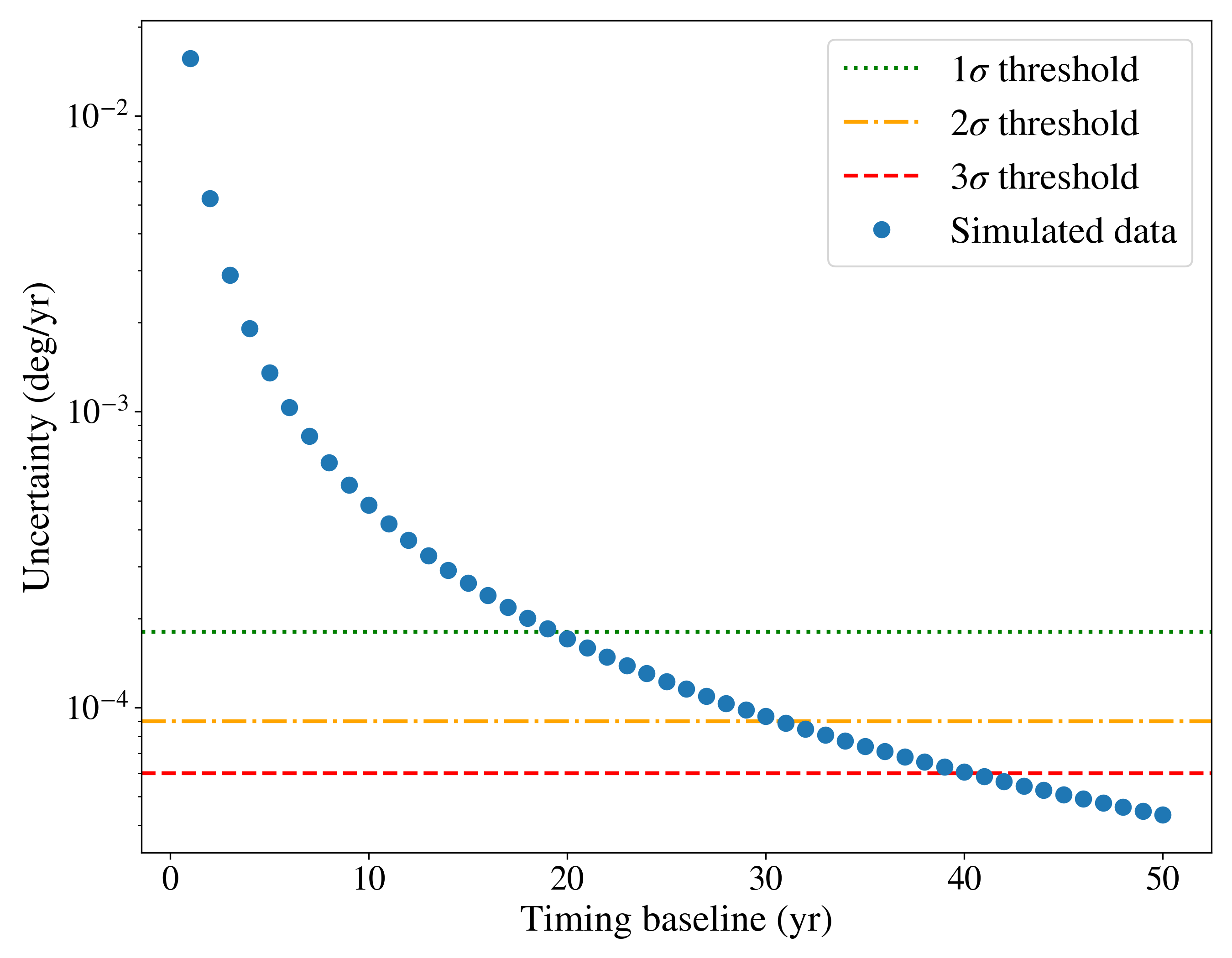} 
    \caption{Simulation of the expected uncertainty of the periastron advance rate $\dot{\omega}$ as a function of the total observational baseline for PSR~J2338+4818. The simulation assumes a typical FAST observation cadence of 14~days and a TOA precision of $30\,\mu\mathrm{s}$. Each blue point (Simulated data) in the figure corresponds to a fit including all TOAs accumulated up to that observational baseline. The dotted green, dash-dotted orange, and dashed red horizontal lines indicate the $1\sigma$, $2\sigma$, and $3\sigma$ detection thresholds, respectively. These thresholds are defined as $\sigma_{\dot{\omega}} = \dot{\omega}_{\mathrm{theory}}/n$ (where $n=1, 2, 3$), with $\dot{\omega}_{\mathrm{theory}} = 0.00018~{\rm deg\,yr^{-1}}$ representing the theoretically predicted value.} 
    \label{fig:periastron_advance_prediction}  
\end{figure}

Owing to the limited timing baseline, the rate of periastron advance cannot be directly determined from our observations.
The theoretically predicted rate of periastron advance is given by \citep{2016ARA&A..54..401O}:
\begin{equation}
    \dot{\omega} = 3 \left( \frac{2\pi}{P_{\rm b}} \right)^{5/3} 
    \frac{T_\odot^{2/3}(M_\mathrm{p}+M_\mathrm{c})^{2/3}}{1-e^2}, 
    \label{eq:6}
\end{equation}
where $\dot{\omega}$ is rate of periastron advance, the $P_{\rm b}$ is the orbital period, $T_\odot$ is the solar mass, $M_\mathrm{p}$ is the mass of the pulsar, and $M_\mathrm{c}$ is the mass of the companion, $e$ is the eccentricity. 
For PSR~J2338+4818, the theoretically predicted value of the $\dot{\omega}$ ranges from approximately 0.000179 to $0.000190\,{\rm deg\,yr^{-1}}$, corresponding to a minimum companion mass of $1.03~\rm M_{\odot}$ and a median companion mass of $1.27~\rm M_{\odot}$, respectively, under the assumption that the pulsar mass is $1.35~\rm M_{\odot}$.
We incorporated the theoretically predicted value into the timing solution and refitted all parameters.
No reasonable fitted value could be obtained within the expected range of the $\dot{\omega}$ (0.000179 to $0.000190\,{\rm deg\,yr^{-1}}$).
With the current observational data, the $\dot{\omega}$ cannot be reliably measured.
To estimate the observational baseline required for a precise measurement of $\dot{\omega}$, 
we performed a simulation using the \textsc{PINT} software (see Figure~\ref{fig:periastron_advance_prediction}). 
We generated synthetic TOAs starting from the first observation epoch, 
assuming a typical FAST observation cadence of 14~days and a TOA precision of $30\,\mu\mathrm{s}$. 
To achieve a $3\sigma$ detection (where the measurement uncertainty $\sigma_{\dot{\omega}} \approx 6 \times 10^{-5}\,{\rm deg\,yr^{-1}}$), a total timing baseline of approximately 40~years would be required.

\subsubsection{The Variability of Orbital Period}

We compare the timing parameters of PSR~J2338+4818 obtained in this work with those reported by \citet{2021MNRAS.508..300C}. 
The differences are small and within the expected uncertainties for most parameters.

While most timing parameters are consistent with those reported by \citet{2021MNRAS.508..300C}, 
we note a significant difference in the orbital period $P_b$. 
\citet{2021MNRAS.508..300C} reported $P_b = 95.25536(2)$~days at the timing epoch MJD~58909.654, 
where the number in parentheses denotes the $1\sigma$ uncertainty in the last digit. 
We obtain the $P_b = 95.19294198(4)$~days at the timing epoch MJD~60558.801.

To quantify the difference, we computed $\Delta P_b$ and $\sigma_{\Delta P_b}$:
\begin{align}
\Delta P_b &= |P_{b,{\rm now}} - P_{b,{\rm previous}}| = 0.06242~{\rm days}, \\
\sigma_{\Delta P_b} &= \sqrt{\sigma_{P_{b,{\rm now}}}^2 + \sigma_{P_{b,{\rm previous}}}^2} \approx 2.0 \times 10^{-6}~{\rm days}.
\end{align}
Converting to seconds, $\Delta P_b \approx 5392.9~{\rm s}$ with an uncertainty of $\sim 1.7~{\rm s}$.

Assuming this difference were caused by a secular orbital period derivative $\dot{P}_b$, 
the required value over the observational baseline ($\Delta T  \approx 1.43\times10^8~{\rm s}$) would be
\begin{equation}
\dot{P}_b^{\rm needed} = |\frac{\Delta P_b}{\Delta T}| \approx 3.78\times 10^{-5}~{\rm s\,s^{-1}},
\end{equation}
the propagated uncertainty is
\begin{equation}
\sigma_{\dot{P}_b^{\rm needed}} = \frac{\sigma_{\Delta P_b}}{\Delta T} \approx 1.2\times10^{-8}~{\rm s\,s^{-1}}.
\end{equation}
From our timing solution, we obtain a fitted value of the $\dot{P}_b$
\begin{equation}
\dot{P}_b = (-7 \pm 5)\times 10^{-10}~{\rm s\,s^{-1}}.
\end{equation}
This value is consistent with zero within the uncertainties, 
and is five orders of magnitude smaller than the $\dot{P}_b^{\rm needed}$.
It demonstrates that the discrepancy in $P_b$ is not physical, but instead arises from the dramatically improved precision of our FAST dataset. 
Thanks to the longer timing baseline and larger number of high-precision TOAs, our measurement of $P_b$ is significantly more accurate than the previous result.

\subsubsection{Significant Flux Variability or Nulling}

Previous studies using the Effelsberg telescope reported the presence of long-term nulling in PSR J2338+4818 \citep{2021MNRAS.508..300C}. 
We performed approximately 2.5 years of multi-epoch observations of PSR~J2338+4818 using FAST. 
Our observations show no significant evidence of long-term nulling.
The long-term nulling reported for PSR~J2338+4818 with the Effelsberg telescope may be caused by significant flux variability rather than true nulling.

\begin{figure}
    \centering  
    \includegraphics[width=0.75\textwidth]{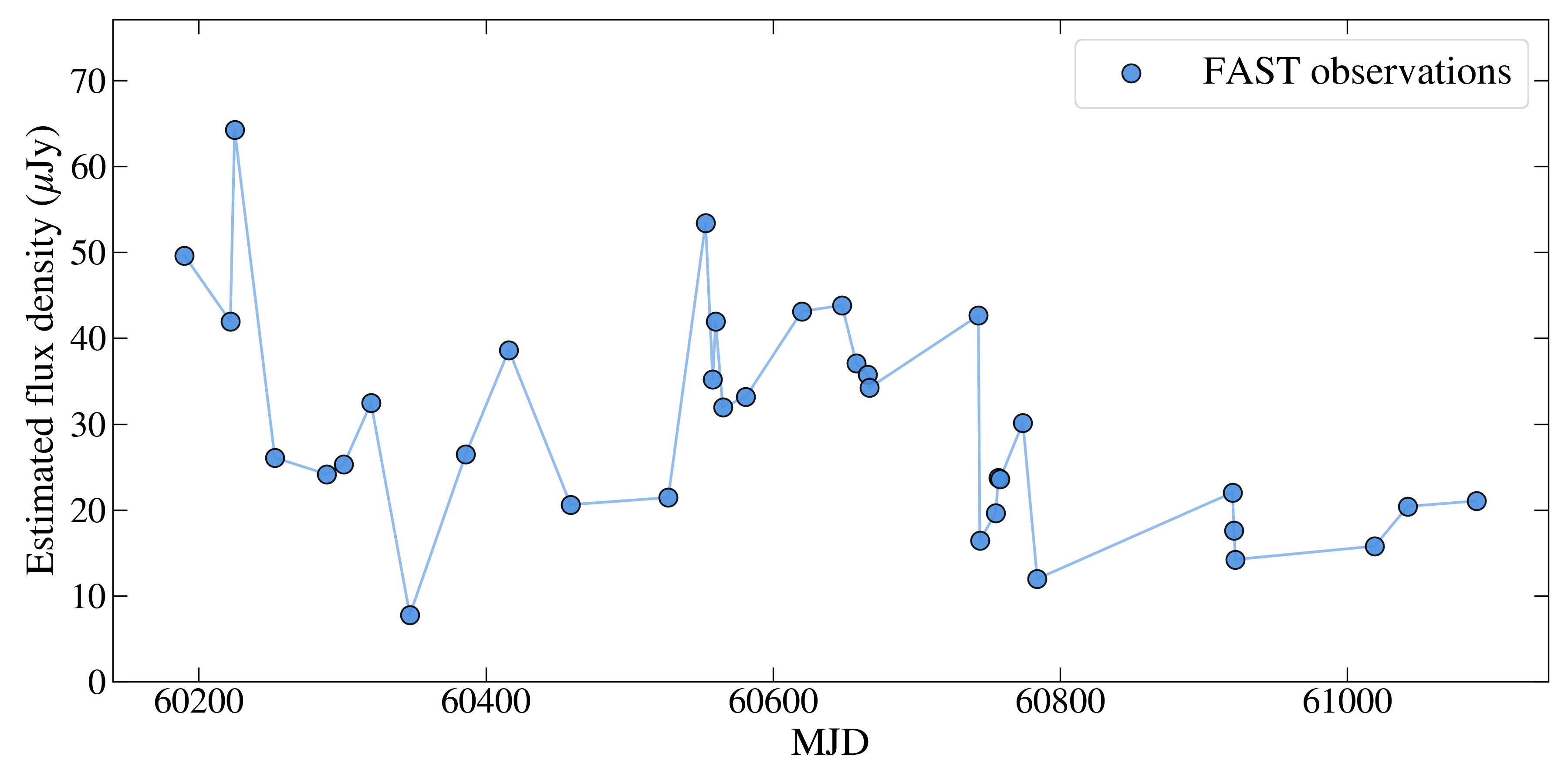}  
        \caption{The estimated flux density of PSR~J2338+4818 across all FAST observations (MJD~60190--61090), excluding the ultra-wideband observation. The flux densities are obtained from the observed S/N and integration time using the equation~\ref{Estimated_flux_density}. Individual observation epochs are shown as data points and are connected by a solid line. Among all observation epochs, the maximum estimated flux density is $64.25~\mu$Jy (MJD~60225), while the minimum is $7.78~\mu$Jy (MJD~60347).} 
    \label{fig:Relative_Flux_Intensity}  
\end{figure}

As shown in Figure~\ref{fig:Relative_Flux_Intensity}, the estimated flux density of PSR~J2338+4818 shows significant variability among the observations, with the ultra-wideband observation excluded.
The maximum value is $64.25~\mu$Jy (MJD~60225), while the minimum is $7.78~\mu$Jy (MJD~60347).
This variability is likely caused by interstellar scintillation.
To determine whether the reported "nulling" is a result of this variability crossing the detection threshold of the Effelsberg telescope, we compared the theoretical sensitivity of FAST and the Effelsberg telescope using the radiometer equation:
\begin{equation}
S_{\text{min}} = \frac{\beta (S/N)_{\text{min}} T_{\text{sys}}}{G \sqrt{n_p \Delta f t_{\text{int}}}},
\label{Estimated_flux_density}
\end{equation}
where $\ \rm S_{\text{min}}$ is the minimum detectable flux density, $\ \rm \beta$ is the system degradation factor due to digitization, $\ \rm (S/N)_{\text{min}}$ is the threshold signal-to-noise ratio, $\ \rm T_{\text{sys}}$ is the system temperature, $\ \rm G$ is the telescope gain, $\ \rm n_p$ is the number of polarizations, $\Delta f$ is the observational bandwidth, and $\ \rm t_{\text{int}}$ is the integration time.
For the FAST observations, the 19-beam L-band receiver parameters are $\ \rm T_{\text{sys}} = 22.0$\,K, $\ \rm G = 16.1$\,K/Jy, and $\Delta f = 400$\,MHz \citep{2020RAA....20...64J}. 
The parameters of the central beam of the 7-beam receiver for Effelsberg observations with the PSRIX pulsar timing backend are $\ \rm T_{\text{sys}} = 23.0$\,K, $\ \rm G = 1.37$\,K/Jy, and $\Delta f = 200$\,MHz \citep{2016MNRAS.458..868L}. 
Assuming the same integration time, FAST is approximately 18 times more sensitive than the Effelsberg telescope. 
Under this sensitivity ratio, a pulse signal detected by FAST with S/N of $\sim 126$ would correspond to S/N of only $\sim 7$ at the Effelsberg telescope. 
For several low-S/N observations in FAST (e.g., MJD~60347 and MJD~60922), where the detected S/N is well below 126, the corresponding signals would fall below the detection threshold of the Effelsberg telescope.

We conclude that the long-term nulling reported in previous literature is likely caused by significant flux density variability due to scintillation and combined with the limited sensitivity of the Effelsberg telescope, rather than true nulling.
Further more observations are required to confirm this interpretation.

\subsubsection{Evolution of Mildly Recycled Pulsars}

PSR~J2338+4818 has a spin period of $P \approx 118$~ms, a period derivative of $\dot{P} \approx 2.3 \times 10^{-18}$, and a small orbital eccentricity. 
In the $P$--$\dot{P}$ diagram, its position lies between the populations of normal pulsars and fully recycled pulsars.
This indicates that PSR~J2338+4818 has undergone only slight recycling and can therefore be classified as a mildly recycled pulsar.

In general, mildly recycled pulsars are typically associated with intermediate-mass binary pulsars (IMBPs) hosting CO or ONeMg WD companions. 
Their relatively slow spins indicate that the neutron star has accreted only a limited amount of mass compared to fully recycled pulsars.
This interpretation is supported by evolutionary studies of recycled pulsars with CO-WD companions, where accretion occurs during Case BB Roche-lobe overflow of a helium star in a post-common-envelope binary, the neutron star typically accretes only $\sim 0.002$--$0.007\, \ \rm M_{\odot}$ \citep{2012MNRAS.425.1601T}.
A comparison with different classes of recycled pulsars further supports this interpretation. 
Recycled pulsars with He-WD companions generally have shorter spin periods, reflecting a larger amount of accreted mass, resulting in fully recycled pulsars.
However, a few binary systems with He-WD companions, such as PSRs~J1232$-$6501, J1810$-$2005, and J1904+0412, exhibit relatively long spin periods and large period derivatives, indicating that they are only mildly recycled \citep{2002ApJ...564..930L}.
These systems have been suggested to originate from IMXBs, where higher mass-transfer rates and shorter evolutionary timescales may limit the total amount of mass accreted by the neutron star.
Whereas binary systems with CO-WD companions typically accrete less mass by a factor of $\sim 6$, leading to less efficient recycling and hence longer spin periods. 
Systems with CO-WD companions are thought to originate from early Case~B or Case~C Roche-lobe overflow, where the mass-transfer phase is relatively short-lived, thereby limiting the total accreted mass and resulting in mildly recycled pulsars with longer $\ \rm P$ and larger $\dot{P}$.
An example is PSR~J2045+3633, which is a mildly recycled pulsar in a $\sim32$ days orbit with CO or ONeMg WD companion \citep{2017MNRAS.470.4421B}. 
Its relatively long spin period and moderate orbital eccentricity ($e \sim 0.017$) are consistent with formation via an IMXB evolutionary channel involving limited mass accretion. 
Compared to this system, PSR~J2338+4818 has a significantly longer orbital period ($P_b \approx 95$~days) and a much lower eccentricity, suggesting likely a wider $P_{b}$ and more circularized orbit of progenitor system.

In contrast, mildly recycled pulsars in DNS systems, which descend from HMXBs, typically exhibit larger orbital eccentricities due to the second supernova explosion. 
Although a small number of DNS systems with lower eccentricities are known ($e \sim 0.06$--$0.1$).
The extremely small eccentricity of PSR~J2338+4818 ($e \sim 0.0018$) indicates that its companion is unlikely to be a neutron star, and instead supports a WD companion formed via stable mass transfer.
A representative example of a mildly recycled pulsar in a DNS system is PSR~J1930-1852. 
This pulsar has a spin period of $P \sim 185$~ms, an orbital period of $P_b \sim 45$~days, and a relatively high eccentricity ($e \sim 0.03$--$0.1$), and its companion is another neutron star \citep{2015ApJ...805..156S}. 
Its moderately short spin period and partial recycling indicate a short or inefficient mass-transfer phase before the companion’s supernova. 
Mildly recycled DNS systems exhibit a positive correlation between spin period and orbital eccentricity \citep{2005MNRAS.363L..71D,2008AIPC..983..464W}. 
The eccentricity of a DNS system is primarily determined by the mass lost during the second supernova explosion and any natal kick imparted to the companion. 
A lower-mass secondary evolves more slowly, prolonging the mass-transfer phase and allowing the primary neutron star to accrete more material, resulting in a shorter spin period and lower eccentricity. 
Conversely, a higher-mass secondary evolves more rapidly, leading to a shorter mass-transfer phase and a longer primary spin period, while mass loss during the supernova produces a higher eccentricity.

\begin{figure}
    \centering  
    \includegraphics[width=0.65\textwidth]{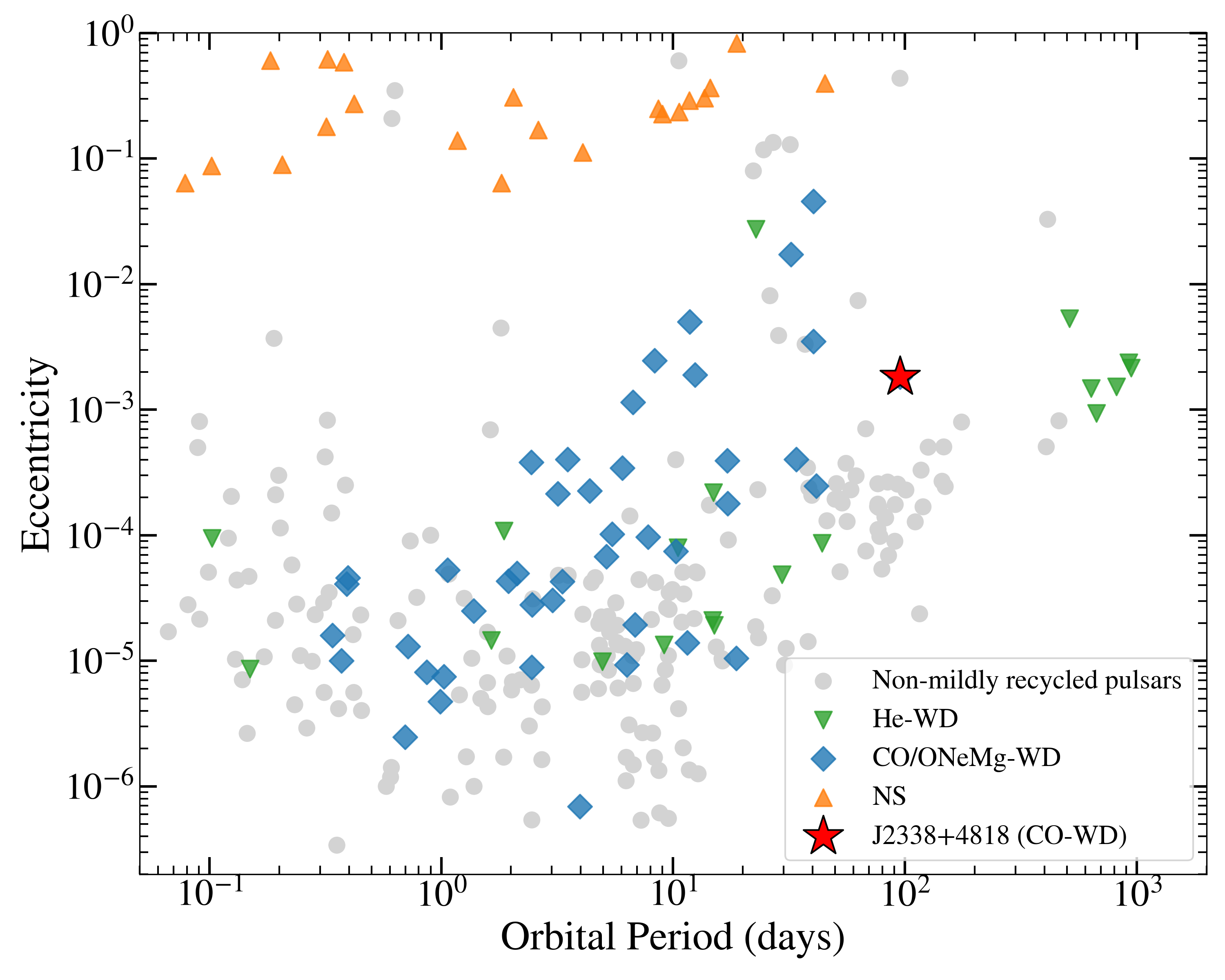}  
        \caption{The $P_b$–$e$ relation for binary pulsars with different types of companions. Binary systems in globular clusters, whose evolution involves companion exchanges or close interactions with other stars, and those with surface magnetic fields $B > 10^{11}$~G, which are thought to have evolved through a different formation channel \citep{2000A&A...355..236T}, both are excluded. Gray circles represent the non-mildly recycled pulsars with spin periods outside the 10--200~ms range. For the mildly recycled pulsars ($10 < P < 200$~ms) in this work, He-WD binary systems are indicated by solid green downward triangles, CO/ONeMg-WD systems by solid blue diamonds, and NS companions by solid orange upward triangles. PSR~J2338+4818 is highlighted with a red star.} 
    \label{fig:PB_ECC}  
\end{figure}

Figure~\ref{fig:PB_ECC} shows the $P_b$–$e$ relation for binary pulsars with different types of companions.
PSR~J2338+4818 exhibits an extremely low eccentricity, which effectively rules out a double neutron star (DNS) origin. 
In constructing this diagram, we exclude binary pulsars in globular clusters, whose evolutionary histories are influenced by dynamical interactions such as companion exchange, as well as systems with magnetic fields $B > 10^{11}$~G, since young pulsars orbiting white dwarfs are thought to have evolved from a different formation channel \citep{2000A&A...355..236T}.
Most CO-WD binary systems cluster at orbital periods $P_b \lesssim 75$~days.
For the mildly recycled pulsars with CO/ONeMg-WD companions, the eccentricity typically increases with the orbital period.
PSR~J2338+4818 extends this population toward longer orbital periods while maintaining a highly circularized orbit.

Overall, the properties of PSR~J2338+4818—its relatively long spin period, extremely low orbital eccentricity, and wide orbit, distinguishes it from most known mildly recycled pulsars with CO white dwarf companions, which typically have $P_b \lesssim 75$~days.
These properties indicate that PSR~J2338+4818 likely experienced a stable mass-transfer phase with limited accretion, and that the progenitor binary system likely had a relatively circular orbit. 
Future multi-wavelength observations, especially optical observations of the companion, may help to determine its physical properties and provide further constraints on the formation and evolution of PSR~J2338+4818.

\subsection{Conclusions}
1. We updated the phase-connected timing solution of the first pulsar discovered by FAST, PSR~J2338+4818, using FAST observations.

2.We have detected single pulses from PSR~J2338+4818, with a total of 27,228 single pulses of $\mathrm{S/N} > 7$.

3. The analysis of the single-pulse energies using a GMM and MCMC sampling indicates that the nulling fraction of this pulsar is negligible. 
We stacked different numbers of pulses with $\mathrm{S/N} < 3$ for the observation at MJD~60743, 
and the results show that a clear pulse signal is still detectable in the average pulse profile.
We find no evidence for significant nulling in our observations.

4. We investigated the scintillation properties of PSR~J2338+4818, 
with distinct scintillation patterns evident in the dynamic spectra. 
We obtained scintillation timescales and bandwidths ranging from 2.93 to 25.26~minutes and 1.68 to 27.41~${\rm MHz}$, respectively.
These results are consistent with those reported by \citet{2021MNRAS.508..300C}. 
No clear scintillation arcs were detected in the secondary spectra across all observations.

5.We demonstrate that the apparent discrepancy in the orbital period $P_b$ of PSR~J2338+4818 is not physical, but instead arises from the limited precision of the earlier timing solution. 
Our simulation indicates that constraining the $\dot{\omega}$ to its theoretically predicted value would require a timing baseline of approximately 40 years.
We further conclude that the long-term nulling reported in earlier studies is likely caused by significant flux density variability due to interstellar scintillation.

\begin{acknowledgements}
This work is supported by the National Key R \& D Program of China No. 2022YFC2205202, No. 2020SKA0120100, NO. 2025SKA0140101 and the National Natural Science Foundation of China (NSFC, Grant Nos. 12373032, 12003047, 11773041, U2031119, 12173052, and 12173053).
Lei Qian is supported by the Youth Innovation Promotion Association of CAS (ID 2018075, Y2022027) and the CAS “Light of West China” Program.
Zhichen Pan is supported by the CAS “Light of West China” Program and the Youth Innovation Promotion Association of the Chinese Academy of Sciences (ID 2023064).
Liyun Zhang is supported by the Science and Technology Program of Guizhou Province under project No. QKHPTRC-ZDSYS[2023]003 and No. QKHFQ[2023]003, and the Guizhou Provincial Natural Science Foundation under project No. ZD[2026]058. 
This work made use of the data from FAST (Five-hundred-meter Aperture Spherical radio Telescope) (https://cstr.cn/31116.02.FAST). 
FAST is a Chinese national mega-science facility, operated by National Astronomical Observatories, Chinese Academy of Sciences.
Finally, we thank the anonymous referee for helpful suggestions to bring clarity to the text.
\end{acknowledgements}

\begin{contribution}

Yujie Chen wrote and revised the manuscript; 
Lei Qian, Zhichen Pan and Liyun Zhang provided guidance on this work design and participated in discussions; the remaining authors contributed to data processing and analysis, and provided support for this work.


\end{contribution}

%
\facilities{FAST}

\software{DSPSR \citep{2011PASA...28....1V},
          PSRCHIVE \citep{2012AR&T....9..237V},
          TransientX \citep{2024A&A...683A.183M},
          PINT \citep{2021ApJ...911...45L,2024ApJ...971..150S},  
          APTB \citep{2024ApJ...964..128T}, 
          Scintools \citep{2020ApJ...904..104R},
          pulsar\_nulling \citep{2023ApJ...948...32A}
          }

\bibliography{apj_ref}{}
\bibliographystyle{aasjournalv7}



\end{document}